\documentclass[apj]{emulateapj}

\shorttitle{Near-IR spectroscopy of a forming cluster at $z=2.16$}
\shortauthors{Tanaka et al.}

\begin{document}

\title{
  On the formation time scale of massive cluster ellipticals\\
  based on deep near-IR spectroscopy at $z\sim2$
}


\author{
  Masayuki Tanaka\altaffilmark{1,2},
  Sune Toft\altaffilmark{3},
  Danilo Marchesini\altaffilmark{4},
  Andrew Zirm\altaffilmark{3},\\
  Carlos De Breuck\altaffilmark{5},
  Tadayuki Kodama\altaffilmark{1},
  Yusei Koyama\altaffilmark{1},
  Jaron Kurk\altaffilmark{6}, and
  Ichi Tanaka\altaffilmark{7}
}
\altaffiltext{1}{National Astronomical Observatory of Japan, Osawa 2-21-1, Mitaka, Tokyo 181-8588, Japan}
\altaffiltext{2}{Kavli Institute for the Physics and Mathematics of the Universe, The University of Tokyo,  5-1-5 Kashiwanoha, Kashiwa-shi, Chiba 277-8583, Japan}
\altaffiltext{3}{Dark Cosmology Centre, Niels Bohr Institute, University of Copenhagen, Juliane Mariesvej 30, DK-2100 Copenhagen, Denmark}
\altaffiltext{4}{Department of Physics and Astronomy, Tufts University, Medford, MA 02155, USA}
\altaffiltext{5}{European Southern Observatory, Karl-Schwarzschild-Str. 2, D-85748 Garching bei M\"{u}nchen, Germany}
\altaffiltext{6}{Max-Planck-Institut f\"{u}r extraterrestrische Physik, Giessenbachstrasse, D-85748 Garching bei M\"{u}nchen, Germany}
\altaffiltext{7}{Subaru Telescope, National Astronomical Observatory of Japan, 650 North A'ohoku Place, Hilo, HI 96720, USA}



\begin{abstract}
We present improved constraints on the formation time scale of massive cluster
galaxies based on rest-frame optical spectra of galaxies in a forming cluster
located at $z=2.16$.
The spectra are obtained with  MOIRCS on the Subaru telescope
with an integration time of $\sim7$ hours.
We achieve accurate redshift measurements 
by fitting SEDs using the spectra and broad-band photometry simultaneously,
allowing us to identify probable cluster members.
Clusters at low redshifts are dominated by quiescent galaxies, but
we find that quiescent galaxies and star forming galaxies co-exist in this $z=2$ system.
Interestingly, the quiescent galaxies form a weak red sequence in
the process of forming.
By stacking the spectra of star forming galaxies, we observe strong emission
lines such as {\sc [oii]} and  {\sc [oiii]} and we obtain a tentative
hint of AGN activities in these galaxies.  
On the other hand, the stacked spectrum of the quiescent galaxies
reveals a clear 4000$\rm\AA$ break with a possible CaII H+K absorption feature
and strong emission lines such as {\sc [oii]} are absent in the spectrum,
confirming the quiescent nature of these galaxies.
We then perform detailed spectral analyses of the stacked spectrum, which 
suggest that these massive quiescent galaxies formed at redshifts between
3 and 4 on a time scale of $\lesssim0.5$ Gyr.
This short formation time scale is not reproduced in recent numerical simulations.
We discuss possible mechanisms for how these galaxies form $10^{11}\rm\ M_\odot$
stellar mass on a short time scale and become red and quiescent by $z=2$.
\end{abstract}


\section{Introduction}

Massive early-type galaxies in the local universe have long been known
to show surprisingly tight scaling relations such as 
Faber-Jackson relation \citep{faber76}, or more generally,
fundamental plane \citep{djorgovski87,jorgensen96}, as well as
red sequence \citep{baum59}.
These observations suggest that early-type galaxies are a fairly
homogeneous population and  they are thought to form in an intense
starburst in their initial phase, followed by passive evolution.
The seminal work of the galactic-wind model by \citet{arimoto87}
is among the first successful theoretical attempts to
reproduce the observed properties of early-type galaxies in
the local universe such as the mass-metallicity relation.

This simple picture of early-type galaxy formation is supported by
the observation that the single-burst passive evolution model is
able to reproduce the observed location of the red sequence even in
$z\sim1$ clusters (e.g., \citealt{stanford98,mei09}).
Detailed absorption line studies of nearby early-type galaxies
lends a further support.  Alpha elements are primarily released in
type-II supernova explosions, which occur on a short time scale.
On the other hand, iron-peaked elements are primarily produced
by type-Ia supernovae, which are thought to occur with a delay
time of $\sim1$ Gyr. 
The observed [$\alpha$/Fe] enhancement at high mass (e.g., \citealt{nelan05})
is often interpreted as massive galaxies being formed on a fairly
short time scale at high redshifts \citep{thomas05} so that type-Ia
supernovae do not significantly contribute to the overall
metal enrichment, although  [$\alpha$/Fe] is also sensitive to IMF.

This picture nicely fits within
the framework of a top-down galaxy formation scenario.
However, the widely accepted $\Lambda$CDM cosmology naturally
predicts bottom-up galaxy formation.  This bottom-up scenario
might appear to confront with the simple observational picture
above, but \citet{delucia06} presented a semi-analytic model of galaxy
formation built upon the Millennium simulation \citep{springel05a}
and showed that the hierarchical model can reproduce the
observed old stellar population of massive early-type galaxies today.
In their model, galaxies do assemble hierarchically, but most stars that
are in massive galaxies at $z=0$ formed early and the stellar
population is thus old.  An extreme example of this hierarchical
assembly  presented by
\citet{delucia07} is that the most massive galaxies today has
acquired 80\% of their mass below redshift of unity.
Deep imaging observations of nearby elliptical galaxies often
unveil tidal features around them, which support the picture
of growing early-type galaxies at low redshifts
(e.g., \citealt{vandokkum05,tal09}).
We are starting to understand the galaxy formation in the hierarchical
universe, but theoretical models in this context still do not fully
reproduce the observed mass growth of galaxies (e.g., \citealt{guo11})
and the massive galaxy formation still remains one of the major issues.

Recent advent of sensitive near-IR spectrographs on large aperture
telescopes has opened a new window in this area.
Imaging observations have shown that many of the massive
galaxies in the field at $z\sim2$ are actively forming stars, but
quiescent galaxies do exist (e.g.,
\citealt{cimatti04,forsterschreiber04,daddi05,williams09,brammer11,wuyts11}).
Deep near-IR spectroscopy of these quiescent galaxies has revealed
their evolved stellar populations.
\citet{kriek06} performed near-IR spectroscopy of $z\sim2$ galaxies
in order to probe their rest-frame optical light and showed that
roughly half of the galaxies they observed have suppressed star
formation.  Ultradeep spectroscopy of
a $z\sim2$ galaxy by \citet{kriek08} showed a pronounced 4000\AA\ break.
\citet{gobat12} and \citet{onodera12} also presented deep near-IR
spectroscopy, and
recently, X-shooter has started producing high-quality spectra of
$1.5\lesssim z\lesssim2$ galaxies with clear detections of
absorption features (e.g., \citealt{toft12,vandesande12}) and
the number of such spectra is fast growing.

So far, most of the work on $z\sim2$ massive galaxies are based on
data in blank fields and the nature of massive cluster galaxies
remains unclear.  But, high-$z$ clusters are an interesting site
to study the formation of massive ellipticals given the prominent
red sequence in local clusters. 
Also, they will provide a key to understanding the origin of
the environmental dependence of galaxy properties observed
at low redshifts (e.g., \citealt{lewis02,gomez03,tanaka04}).
Dedicated effort has identified
a number of high redshift groups and clusters of galaxies at $z\gtrsim1.5$
(e.g., \citealt{kurk09,tanaka10b,fassbender11,gobat11,nastasi11,santos11,spitler11,stanford12,muzzin13,tanaka13}).
Among them, a $z=1.61$ group, which is likely a progenitor of a today's
massive cluster, is carefully studied by \citet{tanaka13} and
they showed that the group already exhibits a prominent red sequence
of quiescent early-type galaxies formed at $z_f=3$ (SSP-equivalent).
This adds further evidence that at least a fraction of today's massive
cluster ellipticals passively evolve since an early epoch.
Higher redshift galaxies have stronger constraining power on
their own formation process, and in this paper, we focus
on a forming cluster at $z=2.16$.
We refer the reader to \citet{strazzullo13} and \citet{gobat13}
for recent work on a $z=2.0$ cluster.

PKS1138-26 is a powerful radio galaxy located at $z=2.16$ and
it has been intensively studied by many authors, providing robust
evidence for a forming cluster
\citep{pentericci97,pentericci98,kurk00,pentericci02,kurk04a,kurk04b,croft05,miley06,kodama07,zirm08,hatch08,hatch09,doherty10,tanaka10,koyama13}.
A wealth of imaging data as well as a large number of spectroscopic
redshifts are available in this field, making it an unique place
to investigate the question raised above.  We present  deep near-IR
spectroscopy of galaxies in this proto-cluster and study galaxy
populations in detail with an emphasis on the formation of
massive cluster ellipticals.

The paper is organized as follows.  We summarize the data used in this work
in section 2.  We perform SED fitting using spectroscopic and photometric
data in this paper and the procedure is described in section 3,
followed by the results from the SED fitting in section 4.
We  discuss the nature of star forming members by stacking their
spectra in section 5.
We then turn our attention to massive quiescent members and perform
spectral analyses to constrain the stellar population and formation
time scale of these massive galaxies using a stacked 
spectrum in section 6.  We summarize and discuss the implications of
our findings in section 7.
Throughout the paper, we assume a flat universe with $\Omega_M=0.27$
and $\Omega_\Lambda=0.73$ with $\rm H_0=70\ km\ s^{-1}\ Mpc^{-1}$ \citep{komatsu11}.
Magnitudes are in the AB system.

\section{Data}

\subsection{Multi-wavelength catalog of PKS1138}

A deep multi-wavelength data set is available in the field
of the radio galaxy PKS1138 at $z=2.16$.  The field has been
imaged with LRIS on Keck ($U$-band; \citealt{zirm08}), FORS
on VLT ($R$ and $z$), ACS on HST ($g$ and $I$; \citealt{miley06}),
MOIRCS on Subaru ($J$ and $K_s$; \citealt{kodama07}),
SOFI on NTT ($H$; \citealt{kodama07}), and IRAC on Spitzer
($3.6-8.0\mu m$; \citealt{seymour07}).
We base our analysis on an updated version of the multi-band catalog
presented in \citet{tanaka10}.  We primarily used {\sc MAG\_AUTO}
from Source Extractor \citep{bertin96} in the original catalog, but
this resulted in somewhat erroneous photometry particularly
for faint objects.  We instead smooth all the images to the worst
seeing of 1.2 arcsec and perform aperture photometry in dual
image mode to derive colors.
All the objects are selected in the unsmoothed $K_s$-band.
Aperture correction is derived by
comparing {\sc MAG\_AUTO} and aperture magnitudes in the $K_s$
band and has been applied to all the other bands.

\subsection{MOIRCS Observation and Data Reduction}

We used MOIRCS on the Subaru Telescope \citep{ichikawa06,suzuki08}
to perform a near-IR follow-up spectroscopy of the galaxies in PKS1138.
The targets for spectroscopy were selected by photometric
redshifts ($z_{phot}$ hereafter) in \citet{tanaka10}. We first applied
a $K_s$-band magnitude cut of $K_s<23$ to ensure that we do
not observe spurious objects.  Galaxies that are consistent with
being at the cluster redshift within $2\sigma$ were selected
as primary targets for spectroscopy.   We further gave
a priority to red objects in the $J-K_S$ color because their $z_{phot}$
are likely more precise and 
we are particularly interested in constraining the formation history of
the oldest, most massive cluster galaxies.
In areas of the MOIRCS field with no suitable targets
(about 10\% of the field of view),
we filled the mask with random objects, which in the end
serve as a good control sample to estimate our redshift accuracy.
We targeted 38 objects in total, of which 34 are photo-$z$ selected
galaxies.  Among them, 13 have a red color with $J-K_S>1$.
Only this one mask was observed in our run.

The observations were carried out on the 4th and 8th Feb. 2011.
The conditions were photometric and seeing was good
(0.6-0.8 arcsec) on both nights. We used $0''.8$ slits
with the zJ500 grism, which provides a wavelength coverage of
0.9-1.7$\mu m$ with a resolving power of $R\sim500$.
The data reduction was performed in a standard manner using
the custom-designed code described in \citet{tanaka09}.
After the dark subtraction, the frames were cut into individual
slits and flat-fielding was performed.  
We measured relative spatial offsets between the exposures
using a bright star in the mask.
As most of the objects were not readily visible in each exposure,
we first stacked all the exposures to measure a trace of an object.
We then extracted objects in each exposure using that trace 
shifted by the spatial offsets measured off the bright star.
Finally, the extracted spectra were combined with weights computed
with the flux of the bright star, which effectively
accounted for both seeing and atmospheric transparency variations.
The wavelength was calibrated against the sky lines and the flux was
calibrated against A0V stars.  Telluric absorptions were corrected for
in each exposure using a bright star observed in the same mask.
We note that the spectral fluxes are not calibrated in the absolute
sense at this point due to the slit loss.
The total science integration time amounted to 6.8 hours.

Among 38 objects that we  observed, we were able to extract
spectra for 30 of them and we have visually inspected all of them.
Even with the long integration on an 8m telescope,
a typical S/N of the spectra is too low ($\sim1.5$ per
resolution element) to
directly measure redshifts via absorption features.
Seven objects show multiple emission lines and their redshifts can
be derived.  Five objects show only a single emission line
and the line is not uniquely identified.
Such a line is probably a strong line such
as {\sc [oii]$\lambda3726/\lambda3729$}, {\sc [oiii]$\lambda5007$} or
H$\alpha\lambda6563$.
We choose the most probable line among them using spectrophotometric
redshifts measured in Section 3.
We cannot measure secure redshifts for the rest of the objects.
We present a list of measured redshifts in Table \ref{tab:spec_sample}
along with spectroscopic redshifts from the literature where 
available \citep{pentericci00,kurk04a,croft05,doherty10}.
Note that our redshift for ID=537 is inconsistent with the literature
redshift, but our redshift is not a secure redshift.
The spectra with measured redshifts (both secure and possible redshifts)
are presented in Appendix.

\begin{table*}
  \begin{center}
    \begin{tabular}{llllllllllll}
      ID & R.A. & Dec. & $K_s$ & $z_{spec}$ & $z_{flag}$ & $z_{pub}$ & $z_{specphot}$ & $P_{cl}$ & $M_{stellar}$ & SFR & $\tau_V$\\\hline
      60 & $11^h40^m53^s.9$ & $-26^\circ 27'21".9$ & 22.42 & 1.175 & 0 & --- & $1.02^{+0.17}_{-0.10}$ & $0.00$ & $0.4^{+0.2}_{-0.1}$ & $0.6^{+0.3}_{-0.3}$ & $0.2^{+0.1}_{-0.2}$\\
      61 & $11^h40^m43^s.3$ & $-26^\circ 27'22".2$ & 21.84 & 0.832 & 0 & --- & $0.67^{+0.05}_{-0.30}$ & $0.00$ & $0.2^{+0.0}_{-0.1}$ & $0.4^{+0.5}_{-0.2}$ & $0.1^{+0.1}_{-0.1}$\\
      63 & $11^h40^m50^s.9$ & $-26^\circ 27'25".4$ & 19.87 & --- & --- & --- & $0.90^{+0.04}_{-0.07}$ & $0.00$ & $8.9^{+2.3}_{-1.8}$ & $0.1^{+0.1}_{-0.0}$ & $0.0^{+0.1}_{+0.0}$\\
      65 & $11^h40^m44^s.1$ & $-26^\circ 27'22".7$ & 21.59 & --- & --- & --- & $1.90^{+0.13}_{-0.17}$ & $0.12$ & $8.4^{+2.8}_{-2.1}$ & $0.3^{+0.9}_{-0.2}$ & $0.3^{+0.1}_{-0.2}$\\
      113 & $11^h40^m52^s.7$ & $-26^\circ 27'30".4$ & 22.28 & 2.348 & 0 & --- & $2.24^{+0.05}_{-0.07}$ & $0.58$ & $1.8^{+0.2}_{-0.3}$ & $1.7^{+1.0}_{-0.2}$ & $0.2^{+0.1}_{-0.1}$\\
      172 & $11^h40^m45^s.3$ & $-26^\circ 27'42".1$ & 21.87 & --- & --- & --- & $1.75^{+0.23}_{-0.21}$ & $0.00$ & $6.7^{+1.3}_{-1.4}$ & $0.1^{+0.1}_{-0.0}$ & $0.1^{+0.2}_{-0.1}$\\
      206 & $11^h40^m49^s.4$ & $-26^\circ 27'50".0$ & 22.67 & 0.842 & 9 & --- & $0.85^{+0.07}_{-0.05}$ & $0.00$ & $0.2^{+0.0}_{-0.0}$ & $1.0^{+0.5}_{-0.4}$ & $0.1^{+0.1}_{-0.1}$\\
      280 & $11^h40^m39^s.2$ & $-26^\circ 28'09".3$ & 22.15 & 2.062 & 9 & --- & $2.13^{+0.04}_{-0.05}$ & $0.85$ & $1.7^{+0.2}_{-0.1}$ & $42.7^{+29.8}_{-32.4}$ & $0.6^{+0.3}_{-0.3}$\\
      286 & $11^h40^m45^s.2$ & $-26^\circ 28'11".0$ & 21.95 & --- & --- & --- & $2.20^{+0.05}_{-0.05}$ & $0.86$ & $2.1^{+0.1}_{-0.2}$ & $13.2^{+38.1}_{-1.4}$ & $0.3^{+0.2}_{-0.1}$\\
      312 & $11^h40^m40^s.0$ & $-26^\circ 28'17".7$ & 22.85 & 1.568 & 9 & --- & $1.50^{+0.27}_{-0.22}$ & $0.00$ & $0.8^{+0.4}_{-0.2}$ & $2.4^{+5.7}_{-1.4}$ & $0.4^{+0.4}_{-0.2}$\\
      329 & $11^h40^m50^s.0$ & $-26^\circ 28'22".2$ & 22.76 & 1.667 & 0 & --- & $1.47^{+0.35}_{-0.13}$ & $0.05$ & $0.6^{+0.4}_{-0.1}$ & $3.2^{+3.5}_{-1.5}$ & $0.3^{+0.1}_{-0.2}$\\
      333 & $11^h40^m54^s.5$ & $-26^\circ 28'23".7$ & 21.34 & --- & --- & --- & $2.07^{+0.05}_{-0.10}$ & $0.52^*$ & $5.6^{+0.7}_{-0.6}$ & $38.0^{+13.3}_{-6.4}$ & $0.9^{+0.1}_{-0.1}$\\
      399 & $11^h40^m39^s.7$ & $-26^\circ 28'45".1$ & 20.96 & --- & --- & 2.162 & $2.22^{+0.06}_{-0.05}$ & $0.70^*$ & $12.6^{+1.5}_{-1.4}$ & $49.0^{+25.2}_{-6.3}$ & $0.8^{+0.1}_{-0.1}$\\
      435 & $11^h40^m49^s.1$ & $-26^\circ 28'52".6$ & 22.33 & 2.037 & 9 & --- & $2.20^{+0.10}_{-0.09}$ & $0.54$ & $2.5^{+0.5}_{-0.3}$ & $104.7^{+18.3}_{-64.0}$ & $1.2^{+0.2}_{-0.2}$\\
      443 & $11^h40^m46^s.1$ & $-26^\circ 28'54".2$ & 21.13 & 1.615 & 0 & --- & $1.57^{+0.04}_{-0.02}$ & $0.00$ & $3.4^{+0.2}_{-0.2}$ & $1.7^{+0.1}_{-0.1}$ & $0.1^{+0.1}_{-0.0}$\\
      445 & $11^h40^m49^s.7$ & $-26^\circ 28'54".5$ & 21.68 & --- & --- & --- & $2.05^{+0.20}_{-0.15}$ & $0.32$ & $8.9^{+1.7}_{-1.8}$ & $0.1^{+0.2}_{-0.1}$ & $0.1^{+0.2}_{-0.1}$\\
      455 & $11^h40^m51^s.4$ & $-26^\circ 28'57".2$ & 20.39 & --- & --- & --- & $2.28^{+0.02}_{-0.04}$ & $0.29$ & $16.8^{+1.0}_{-0.9}$ & $0.9^{+0.1}_{-0.7}$ & $0.2^{+0.1}_{-0.1}$\\
      491 & $11^h40^m44^s.2$ & $-26^\circ 29'07".0$ & 20.78 & 2.162 & 1 & 2.172 & $2.21^{+0.05}_{-0.25}$ & $0.50^*$ & $13.3^{+3.5}_{-0.7}$ & $0.1^{+0.6}_{-0.0}$ & $0.0^{+0.2}_{-0.0}$\\
      493 & $11^h40^m47^s.9$ & $-26^\circ 29'06".3$ & 21.41 & --- & --- & --- & $2.18^{+0.08}_{-0.05}$ & $0.73^*$ & $4.7^{+0.9}_{-0.3}$ & $4.7^{+0.7}_{-0.4}$ & $0.4^{+0.0}_{-0.0}$\\
      534 & $11^h40^m40^s.6$ & $-26^\circ 29'15".3$ & 22.57 & 1.615 & 9 & --- & $1.61^{+0.06}_{-0.10}$ & $0.00$ & $0.8^{+0.0}_{-0.1}$ & $4.7^{+0.6}_{-0.8}$ & $0.3^{+0.0}_{-0.0}$\\
      537 & $11^h40^m45^s.9$ & $-26^\circ 29'16".7$ & 20.44 & 1.335 & 1 & 2.1568 & $0.10^{+0.02}_{-0.02}$ & $0.00^*$ & $0.0^{+0.0}_{-0.0}$ & $0.0^{+0.0}_{-0.0}$ & $0.0^{+0.0}_{+0.0}$\\
      548 & $11^h40^m53^s.7$ & $-26^\circ 29'18".4$ & 21.63 & 2.547 & 0 & --- & $2.87^{+0.05}_{-0.05}$ & $0.00$ & $3.2^{+0.2}_{-0.3}$ & $117.5^{+8.4}_{-10.3}$ & $0.3^{+0.0}_{-0.0}$\\
      559 & $11^h40^m44^s.4$ & $-26^\circ 29'20".7$ & 21.51 & --- & --- & 2.162 & $1.84^{+0.13}_{-0.15}$ & $0.06^*$ & $5.3^{+1.4}_{-1.3}$ & $1.0^{+1.2}_{-0.3}$ & $0.1^{+0.3}_{-0.1}$\\
      582 & $11^h40^m46^s.1$ & $-26^\circ 29'24".8$ & 21.95 & 2.155 & 0 & 2.1546 & $2.18^{+0.07}_{-0.03}$ & $0.84^*$ & $2.1^{+0.1}_{-0.1}$ & $66.1^{+29.4}_{-7.2}$ & $0.6^{+0.2}_{-0.1}$\\
      593 & $11^h40^m46^s.4$ & $-26^\circ 29'26".9$ & 20.86 & --- & --- & --- & $2.09^{+0.05}_{-0.08}$ & $0.61$ & $18.8^{+1.1}_{-3.0}$ & $0.8^{+0.1}_{-0.4}$ & $0.0^{+0.1}_{+0.0}$\\
      635 & $11^h40^m51^s.2$ & $-26^\circ 29'38".5$ & 20.91 & --- & --- & 2.152 & $2.07^{+0.09}_{-0.03}$ & $0.42^*$ & $10.0^{+0.6}_{-0.6}$ & $14.5^{+18.7}_{-1.6}$ & $0.7^{+0.1}_{-0.1}$\\
      669 & $11^h40^m51^s.5$ & $-26^\circ 29'45".6$ & 23.42 & --- & --- & --- & $2.04^{+0.17}_{-0.68}$ & $0.34^*$ & $0.8^{+0.4}_{-0.4}$ & $1.6^{+3.7}_{-1.0}$ & $0.3^{+0.4}_{-0.2}$\\
      705 & $11^h40^m38^s.3$ & $-26^\circ 29'53".5$ & 22.05 & --- & --- & --- & $1.75^{+0.03}_{-0.05}$ & $0.00$ & $5.0^{+0.3}_{-0.5}$ & $0.1^{+0.1}_{-0.1}$ & $0.1^{+0.1}_{-0.1}$\\
      760 & $11^h40^m52^s.6$ & $-26^\circ 30'06".9$ & 22.35 & --- & --- & --- & $1.61^{+0.08}_{-0.29}$ & $0.00$ & $0.8^{+0.3}_{-0.4}$ & $6.5^{+10.5}_{-5.5}$ & $0.4^{+0.3}_{-0.4}$\\
      826 & $11^h40^m39^s.7$ & $-26^\circ 30'22".5$ & 21.93 & --- & --- & --- & $0.24^{+0.06}_{-0.14}$ & $0.00$ & $0.0^{+0.0}_{-0.0}$ & $0.1^{+0.1}_{-0.0}$ & $0.2^{+0.2}_{-0.2}$\\
      \hline
    \end{tabular}
  \end{center}
  \caption{
    Objects observed with MOIRCS.
  }
  \tablecomments{
    The 4th column shows $K_s$ magnitude and the 5th
    and 6th columns are spectroscopic redshift and redshift flag measured from the MOIRCS spectra.
    The flags mean: 0=secure, 1=possible, 9=single emission line.
    For $z_{flag}=9$, the line is identified using $z_{specphot}$.  
    The 7th column is spectroscopic redshifts from the literature \citep{kurk04a,croft05,doherty10}.
    The 8th column is $z_{specphot}$ and the 9th column shows $P_{cl}$ derived
    from $z_{specphot}$ (see Section 4 for the definition of $P_{cl}$).
    Objects with $^*$ on $P_{cl}$ are H$\alpha$ emitters identified by \citet{koyama13}.
    The 10th to 12th columns show physical properties of the galaxies measured from
    the spectrophotometric fits (Section 3).
    Stellar mass is in unit of $10^{10}\rm M_\odot$ and SFR is in $\rm M_\odot\ yr^{-1}$.
    The last column shows the amount of extinction (optical depth in the $V$-band)
    and the conventional $A_V$ can be computed as $A_V=1.09\tau_V$.
  }
  \label{tab:spec_sample}
\end{table*}

\section{Spectral Energy Distribution Fitting}

As mentioned in the last section, a large fraction of
the MOIRCS spectra is not of sufficient quality to
measure redshifts using absorption features.  As a result,
we cannot determine precise redshifts for quiescent galaxies.
However, we can still determine fairly accurate redshifts
as well as physical properties of these galaxies using continuum
features in the spectra such as the 4000$\rm\AA$ break when
combined with the broad-band photometry.
This technique has been demonstrated by
\citet{kriek06} and \citet{kriek08}.
We measure an inverse-variance weighted mean flux of
the observed spectra in 300$\rm \AA$ bins starting from
0.9$\mu m$ to 1.7$\mu m$.
The observed spectra are not flux-calibrated
in the absolute sense due to the slit loss.  We compute
the slit loss by comparing the broad-band $J$ and $H$-band
photometry with the photometry synthesized with the observed
spectra.  We take the average of the slit loss in the
$J$ and $H$-band for most objects.  For a small
number of objects, either $J$ or $H$-band is not fully
covered in the spectra due to the slit positions, and
we use either one of the bands to compute the slit loss.
A typical amount of the slit loss is about 40\%.

We follow the standard procedure for the SED fitting.
We use an updated version of the \citet{bruzual03} code,
which includes an improved treatment of thermally pulsating
AGB stars\footnote{
  Although the details of the improvements have not
  yet been published, the model incorporates a prescription
  of thermally pulsating AGB stars of \cite{marigo07}
  (see section 5.2 of \citealt{eminian08}).
}, to generate model templates of galaxies.
We assume the Chabrier initial mass function and solar
metallicity.  We adopt the $\tau$-model for star formation
histories of galaxies with $\tau$ allowed to vary between
0 and infinity.  Dust extinction is applied to the templates
assuming the Calzetti attenuation curve \citep{calzetti00}.
Emission lines are added to the spectra using the emission
line intensity ratios given in \citet{inoue11} assuming
the \citet{calzetti97} attenuation law.
Each template is convolved with the response functions of
all the filters (including the atmosphere for ground-based facilities).
For the binned spectra, we use a top-hat function to
synthesize fluxes.

The observed SEDs are fit with the model templates using
the standard $\chi^2$ minimization technique.
We apply a template error function in order to reduce
systematics between the model templates and real SEDs of galaxies
and also to assign uncertainties to the model templates
as a function of rest-frame wavelength.
We use the same template error function as in \citet{tanaka13}
(for details, refer to the Appendix B of that paper).
The model templates have 4 free parameters: redshift,
star formation time scale ($\tau$), extinction ($\tau_V$),
and age.  Stellar mass and SFR can be computed from
the combination of redshift, $\tau$, and age.
The $\chi^2$ fits thus produce multi-parameter probability space.
In this probability space, 
each parameters are marginalized over all the other parameters
and we use the median of the probability distribution as the central
value and quote 68\% interval around it as an uncertainty.
Our estimates of the physical parameters thus properly include
uncertainties in redshift.

As mentioned above, we bin the spectra into $\Delta\lambda=300\rm\AA$.
From a statistics point of view, one does not necessarily have to
bin the spectra.  Fitting the unbinned spectra together with
the broad-band photometry may be more straightforward.
We find that, in our case, fits
to unbinned spectra often result in catastrophic outliers
in redshift with unrealistically small uncertainties.
We suspect that this likely comes from non-Gaussian nature of
the noise property of our spectra.  Our spectral resolution is very low
and sky emission lines often overlap each other in near-IR.
It is not surprising if we have under/over subtracted the
sky lines, which would introduce non-Gaussianity in the noise
characteristics of the spectra.
A flux calibration error and small offsets in object traces used for
the 1D extraction, especially at the edges of the covered
wavelength range, are also sources of systematics.
To reduce the systematic errors,
we bin the spectra by clipping outliers in each bin.  To be specific,
we clip top 10\% and bottom 10\% distribution of the fluxes in each
bin and use the rest to compute the weighted average.
This binning at the same time loses wavelength resolution.
We would like to keep the resolution, while having the bin size
wide enough to reduce the systematics.  Our bin size
is a compromise between them.
We have confirmed that our results do not significantly change
if we change the bin size by a factor of 1.5 (i.e., $200\rm\AA$ or $450\rm\AA$).

We denote photometric redshifts based on broad-band data only as
$z_{phot}$ and those based on both broad-band photometry and
binned spectra as $z_{specphot}$ in what follows.
\citet{koo99} suggested to use the term photometric redshifts
for those based on photometry with a spectral resolution of
$\lambda/\Delta \lambda\lesssim20$.
Our binned spectra have a higher resolution and we use
the term $z_{specphot}$ just to distinguish it from $z_{phot}$. 
It will be
instructive to compare redshifts and physical parameters
estimated with and without the binned spectra and we briefly summarize how
the binned spectra improve them in Appendix.

\section{Results}

\subsection{Redshift distribution}

We have performed the SED fitting using the broad-band photometry
and the binned spectra for all of the objects for which we are able
to extract the spectra.  We first investigate how accurate our spectro-photometric
redshifts ($z_{specphot}$) are and then move on to discuss detailed
physical properties of the galaxies in the proto-cluster.

\begin{figure}
\epsscale{1.0}
\plotone{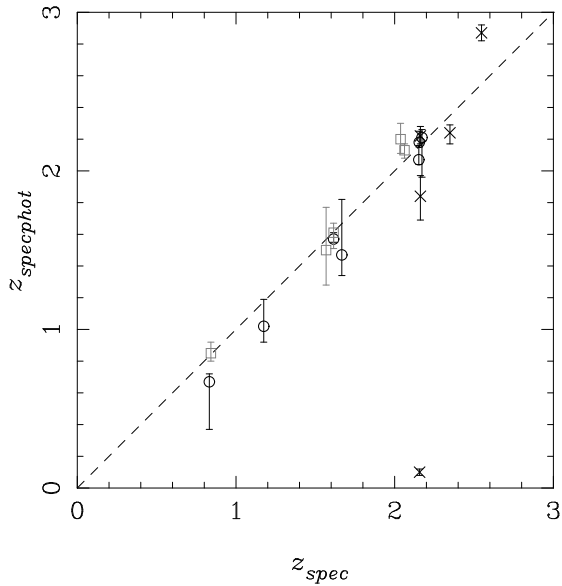}
\caption{
  $z_{specphot}$ plotted against $z_{spec}$.
  The dark open circles are secure redshifts and gray open squares
  are single-line redshifts.  The crosses indicate AGNs.
}
\label{fig:specz_vs_specphotoz}
\end{figure}

Fig. \ref{fig:specz_vs_specphotoz} compares $z_{spec}$ and $z_{specphot}$.
In general, the agreement between them is fairly good, except for AGNs.
This is not surprising because we did not include AGN templates in
the SED fitting. If we do so, $z_{specphot}$ for normal galaxies tends to
degrade and we choose not to include AGN templates.
We measure the accuracy of $z_{specphot}$ to be $\sigma(\Delta z/(1+z))\sim0.03$.
This may not appear as good as one might expect, but this accuracy
is for faint galaxies with $K_s\sim21-22$, while photo-$z$ accuracies
quoted in the literature often include much brighter galaxies.

\begin{figure}
\epsscale{1.0} \plotone{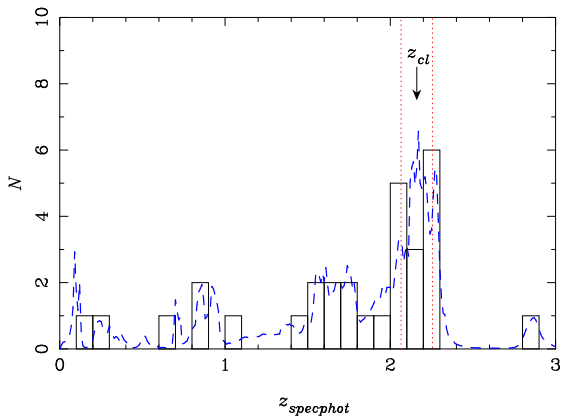}
\caption{
  $z_{specphot}$ distribution. 
  The open histogram shows the redshift distribution of
  galaxies using the central values of $z_{specphot}$, while
  the dashed line shows the full $P(z)$ distribution.
  The cluster redshift is indicated by the arrow.
  The vertical dotted lines show the redshift
  range over which we integrate the PDF in Eq. 1.
}
\label{fig:zhist}
\end{figure}

Fig. \ref{fig:zhist} presents $z_{specphot}$ distribution
of all the objects we observed.  A half of the observed galaxies
are clustered around the cluster redshift of $z=2.16$.
This is particularly clear in the full $P(z)$ distribution and
a sharp peak is observed at the cluster redshift.
The peak is partly due to the fact that we pre-selected member
candidates using $z_{phot}$,
but the $\pm2\sigma$ range in $z_{phot}$ used for the
target selection is typically $\pm0.3$
which is wider than the redshift spike observed here.
This likely represents a large-scale structure and/or
galaxy concentration at the cluster redshift.

We define a proto-cluster membership candidate using the photo-$z$
probability distribution function (PDF).  To be specific, we apply
the following criterion:

\begin{equation}
P_{cl}=\int_{z_{cl}-0.03\times (1+z_{cl})}^{z_{cl}+0.03\times (1+z_{cl})} P(z) dz \geq 0.16,
\end{equation}

\noindent
where $z_{cl}$ is the proto-cluster redshift of $z_{cl}=2.16$ and $P(z)$
is the photo-$z$ PDF.
Note that we take the redshift of the radio galaxy \citep{vanojik95}
as the proto-cluster redshift here.
Our $z_{specphot}$ has $\sigma(\Delta z/(1+z))\sim0.03$
as noted above and 
we adopt the $\pm1\sigma$ range to integrate the PDF.
If the integrated probability exceeds 0.16,
we define the galaxy as a member candidate.  This probability threshold is
arbitrary, but if a galaxy is consistent with being at
$|z_{phot}-z_{cl}|/(1+z_{cl})<0.03$ within $1\sigma$,
it is a candidate for group membership.
Given the limited accuracy of our $z_{specphot}$, we expect some contamination
of near-foreground/background galaxies.  We will estimate the amount of
such contamination later.

Most objects at $2<z_{specphot}<2.3$ satisfies $P_{cl}>0.16$ and
they are the subject of this work.  However, there are three
objects that have $P_{cl}>0.16$ but with spec-$z$'s inconsistent
with being cluster members (ID113, 280, and 435).  We exclude
them from the following analysis.
There are two AGNs with spec-$z$'s consistent with
the cluster members, but we fail to obtain consistent
$z_{specphot}$ due probably to the AGN contamination to
the overall SEDs (ID=537 and 559).
We could fit these objects with redshifts fixed to their
spectroscopic redshifts to derive their physical properties,
but the AGN contribution would still affect the measurement of
SFR and stellar mass.
We do not examine these galaxies either.
This leaves us with 11 member candidates.

\subsection{Comparison with H$\alpha$ emitters}
\label{sec:ha_emitters}

Among the 11 member candidates we have selected, only 2 have spec-$z$'s
from the literature and they are consistent with being the proto-cluster members.
It would be useful to further
check how well we can sample members with $z_{specphot}$.
\citet{koyama13}
performed a narrow-band H$\alpha$ observation of this proto-cluster and
here we cross-match our member candidates with the H$\alpha$ emitters.
We note that their observation reaches to a dust-free SFR of
$\sim6\rm\ M_\odot\ yr^{-1}$ \citep{kennicutt12} and lower SFR galaxies
will not be matched.
Obviously, there are three categories to be looked at;
(a) member candidates and H$\alpha$ emitters, (b) member candidates,
but not H$\alpha$ emitters, and (c) not member candidates, but
H$\alpha$ emitters.

\noindent
{\bf (a):} 7 objects are both member candidates and H$\alpha$
emitters.  They are thus very likely real members.  As we will discuss
below, there are 4 quiescent members and one of them (ID=491) is
detected in H$\alpha$ and falls
in this category.  It may sound surprising that a quiescent galaxy
is detected in H$\alpha$, but this object hosts an AGN \citep{croft05}.
The AGN does not contribute significantly to the overall SED
and this galaxy is included in the analysis.\\
{\bf (b):} 4 objects are member candidates, but not H$\alpha$ emitters.
3 of them are quiescent galaxies with SFR$<1\rm M_\odot\ yr^{-1}$
measured from the SED fitting.
Thus, the non-detection in H$\alpha$ is consistent with that.
The fourth object (ID=286) is a star forming galaxy
with SFR$=13^{+38}_{-1}\rm\ M_\odot\ yr^{-1}$ and $\tau_V=0.3^{+0.2}_{-0.1}$.
Taking into account the extinction, the $1\sigma$ uncertainty in
the expected H$\alpha$ luminosity touches the observational detection limit.
This galaxy may be a near-field contaminant, but it may just have
weak H$\alpha$.\\
{\bf (c):} Two H$\alpha$ emitters are not member candidates,
but they are the AGNs mentioned above for which we fail to
obtain correct $z_{specphot}$.  These objects are excluded from the main analysis.

Overall, our member selection works well.
Of the 11 member candidates with $P_{cl}>0.16$, 7 are detected
in H$\alpha$.  The remaining 4 galaxies are mostly quiescent
galaxies and no H$\alpha$ detections are expected.
This is an encouraging result and it motivates us to study
the physical properties of the member candidates in detail.

\subsection{SEDs of the member candidates}

\begin{figure*}
\epsscale{1}
\plottwo{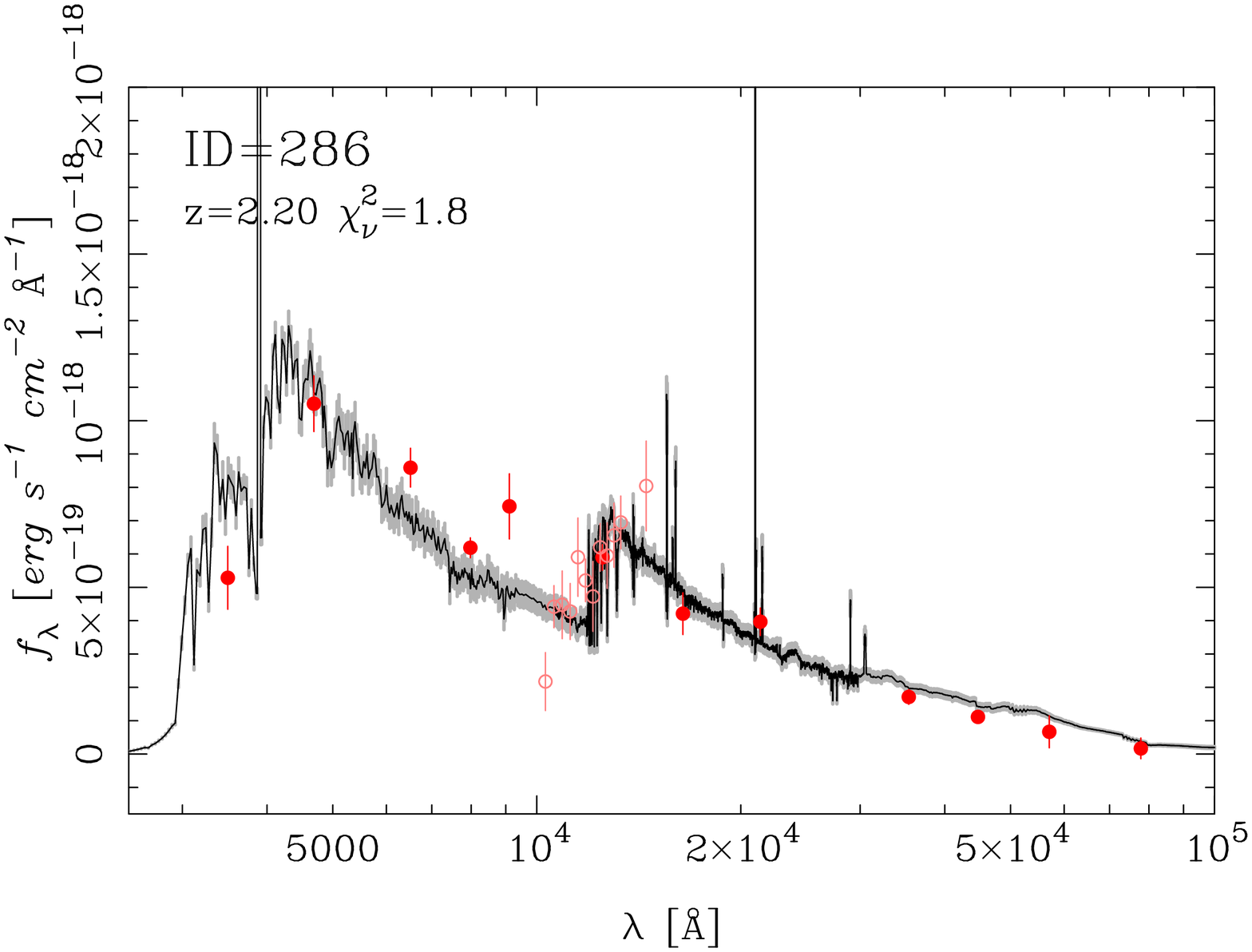}{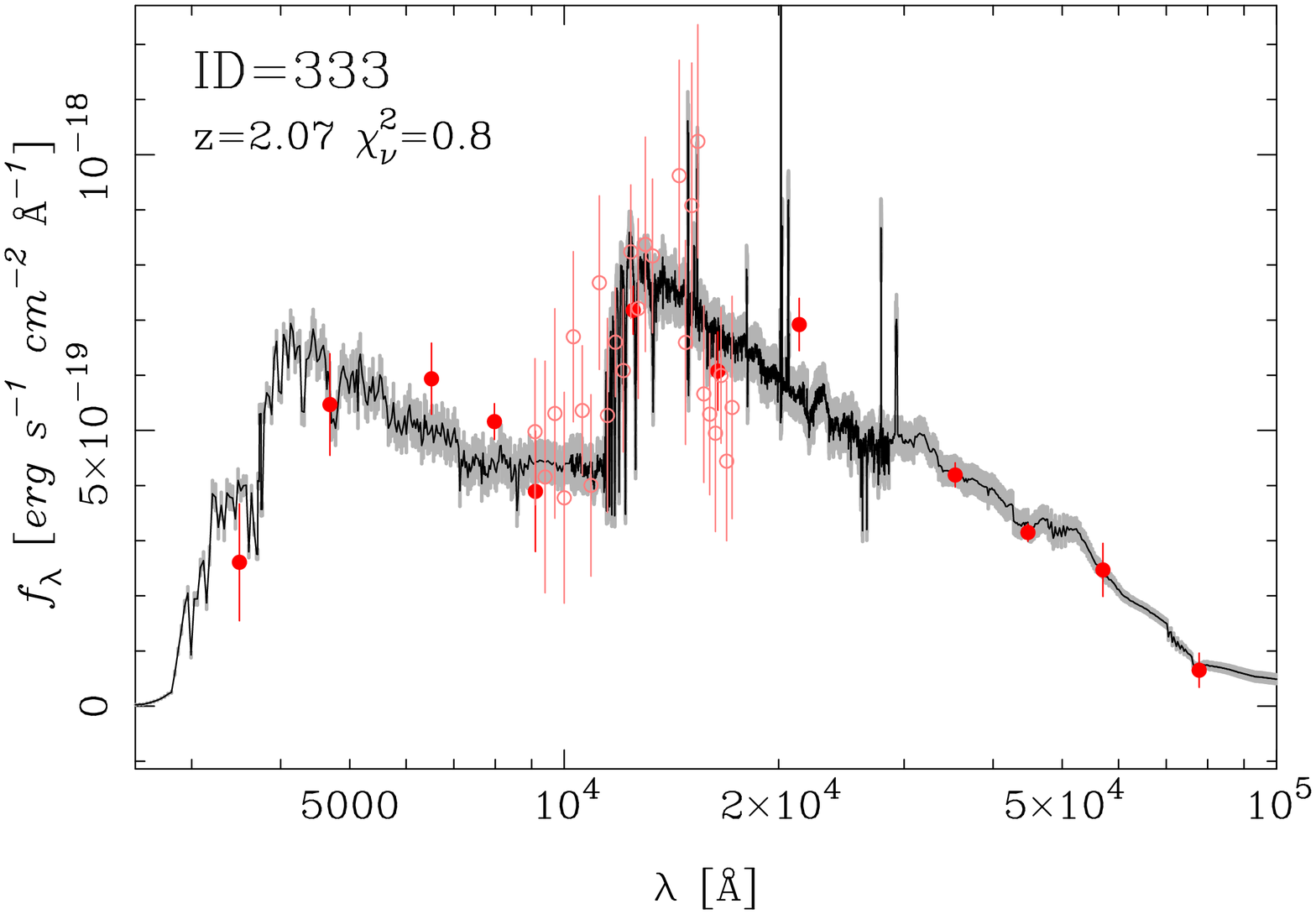}\\
\plottwo{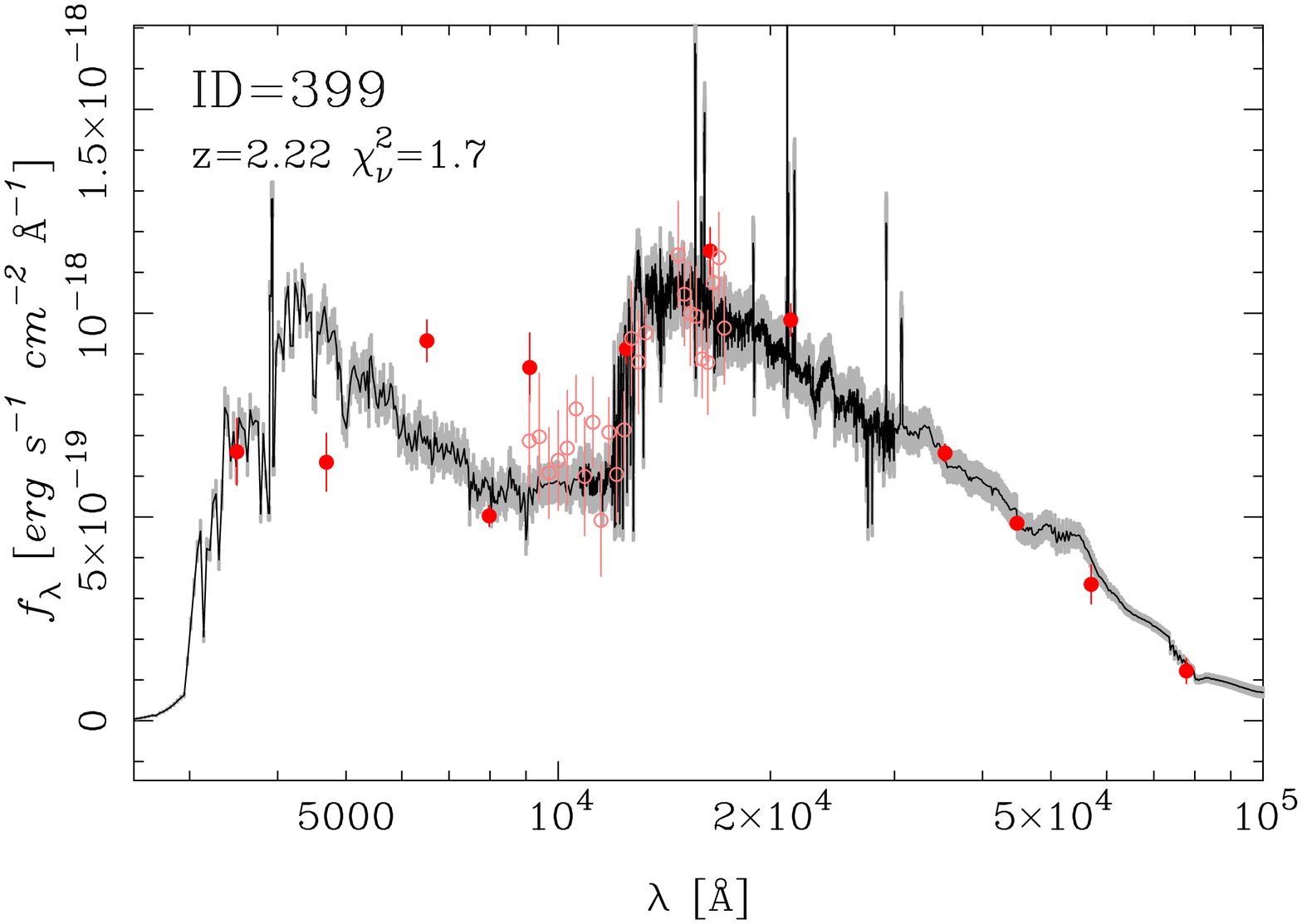}{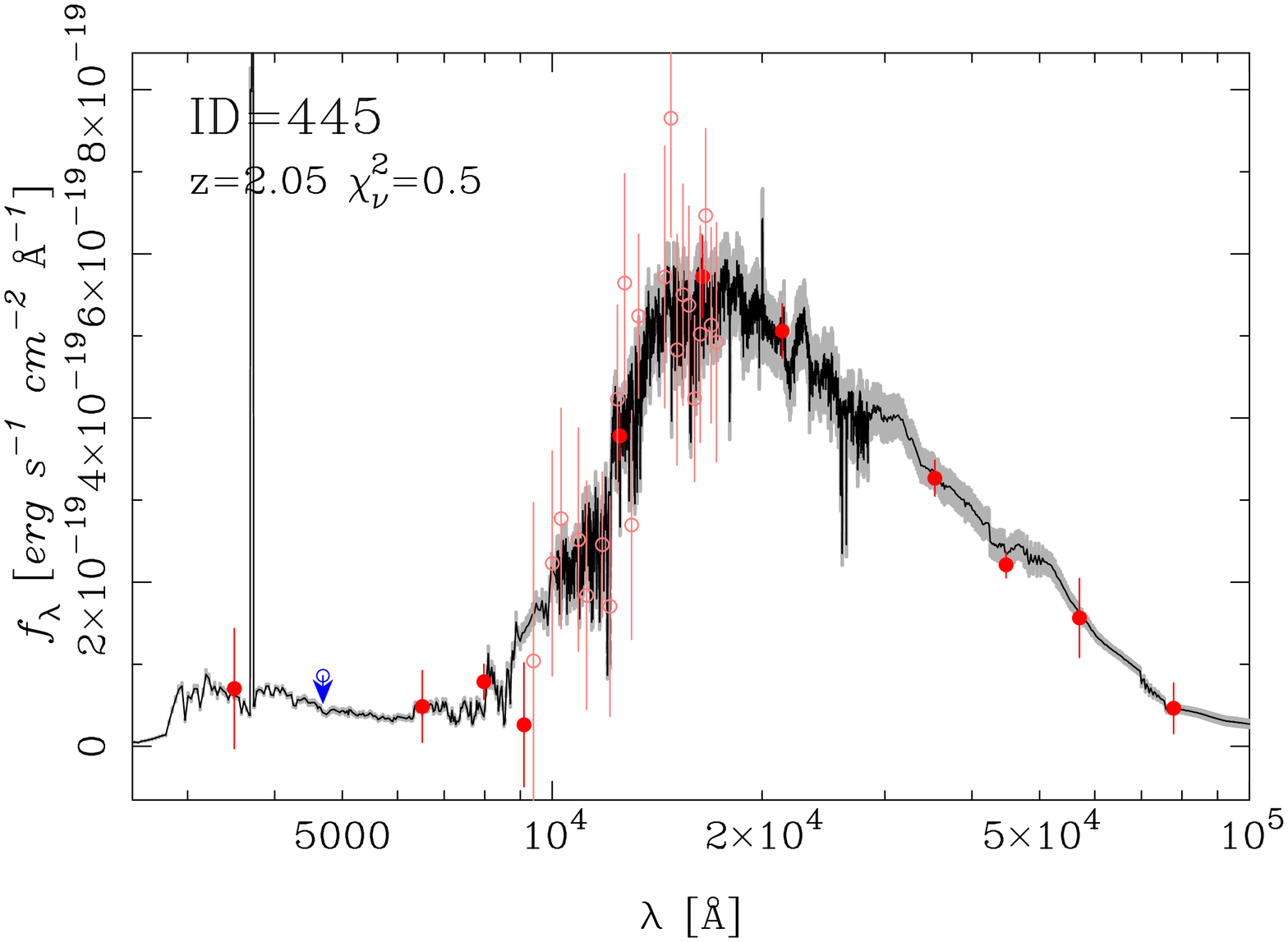}\\
\plottwo{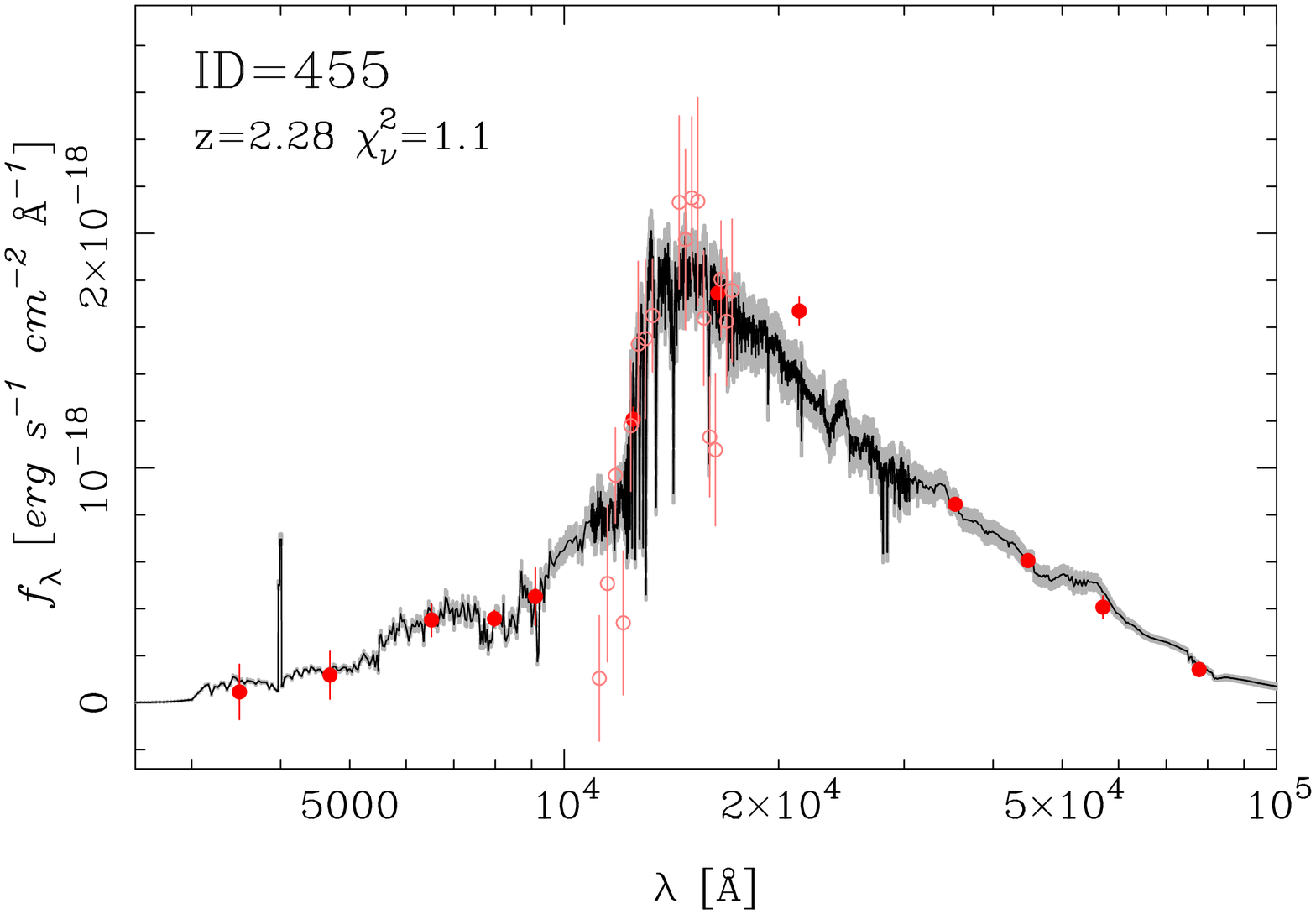}{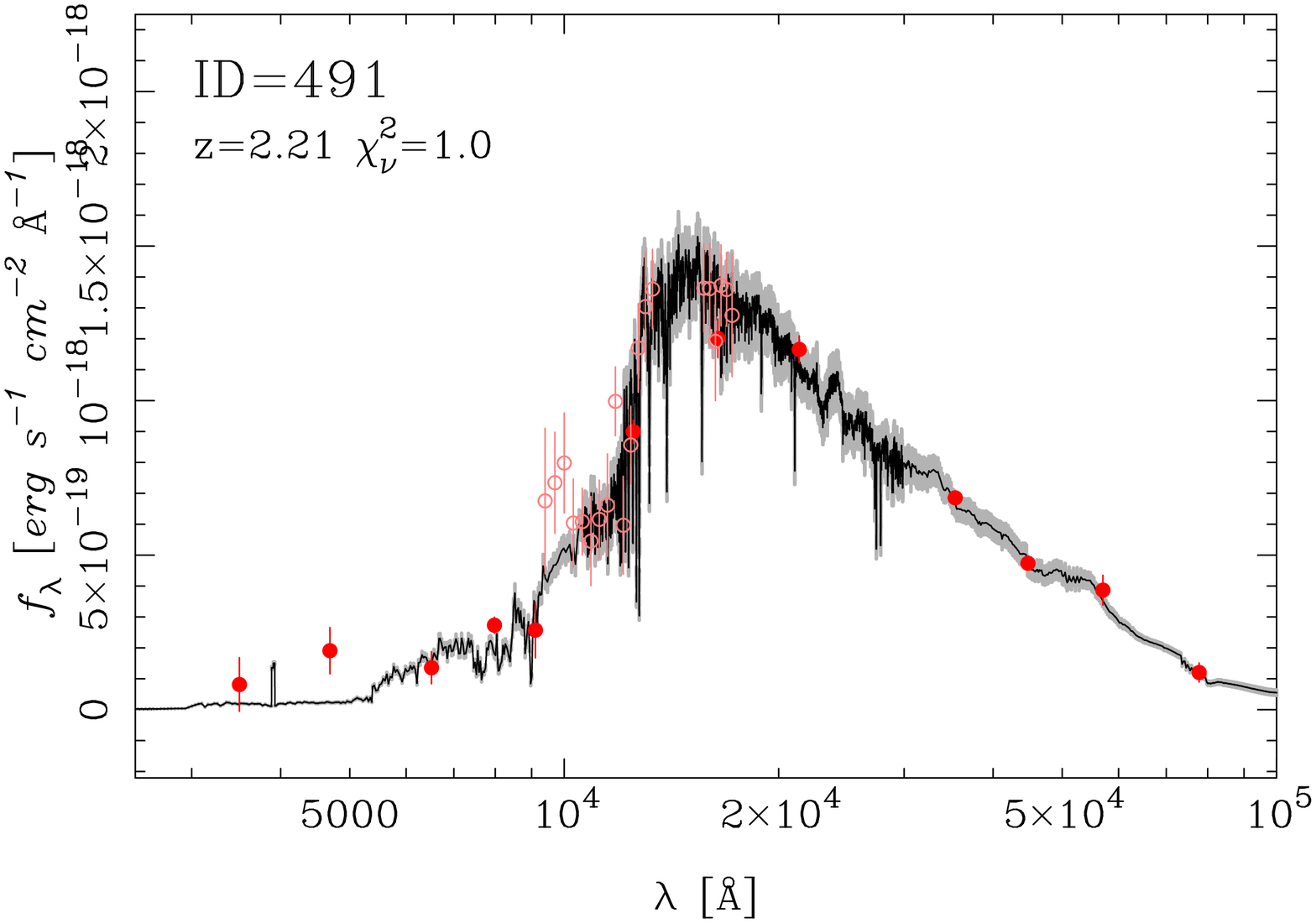}
\caption{
  SEDs of the member candidates.
  The filled and open circles are the broad-band photometry
  and binned spectra, respectively.  The spectrum is
  the best-fitting model template and the shades show
  the uncertainty in the templates.
}
\label{fig:sed1}
\end{figure*}

\begin{figure*}
\epsscale{1}
\plottwo{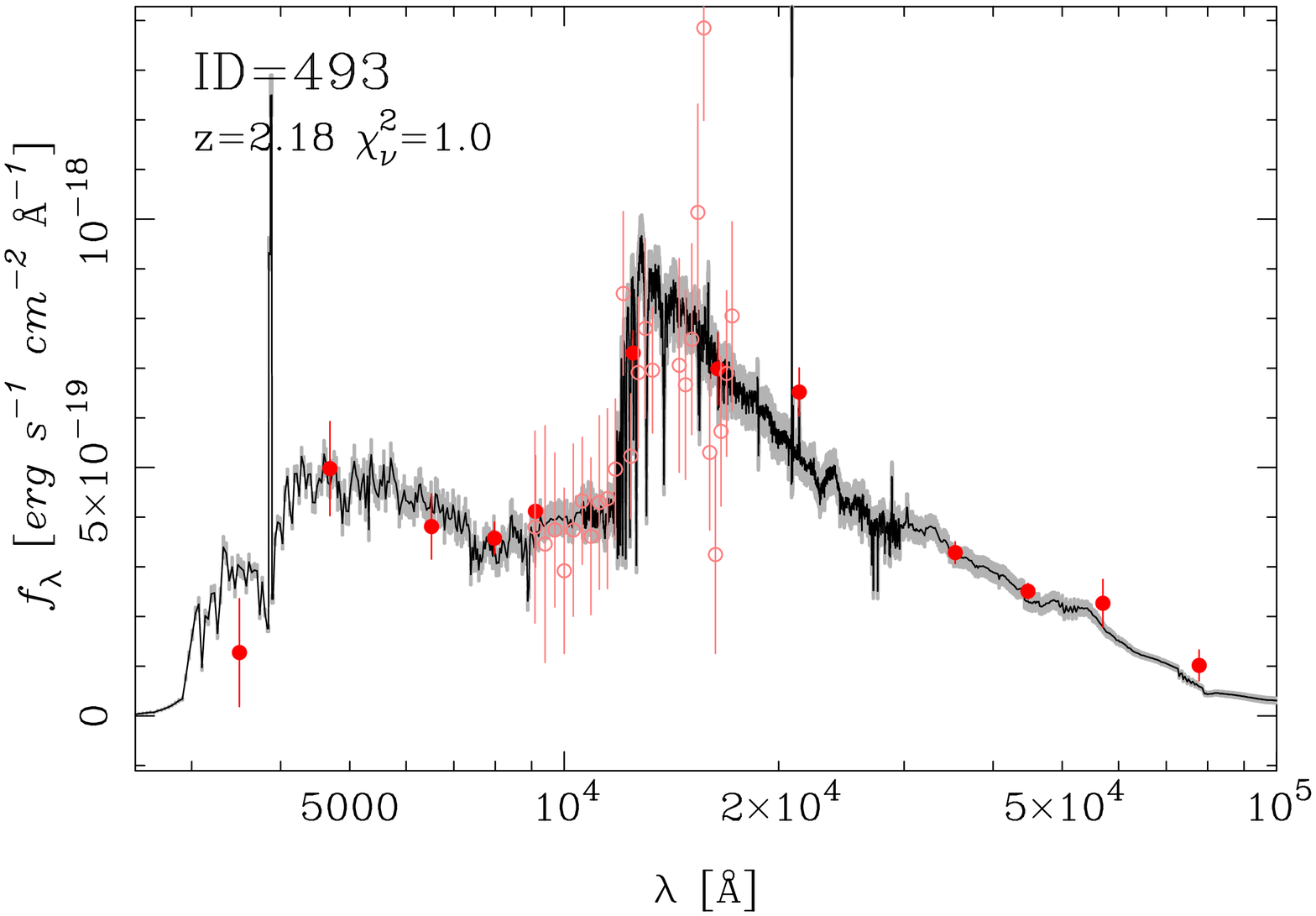}{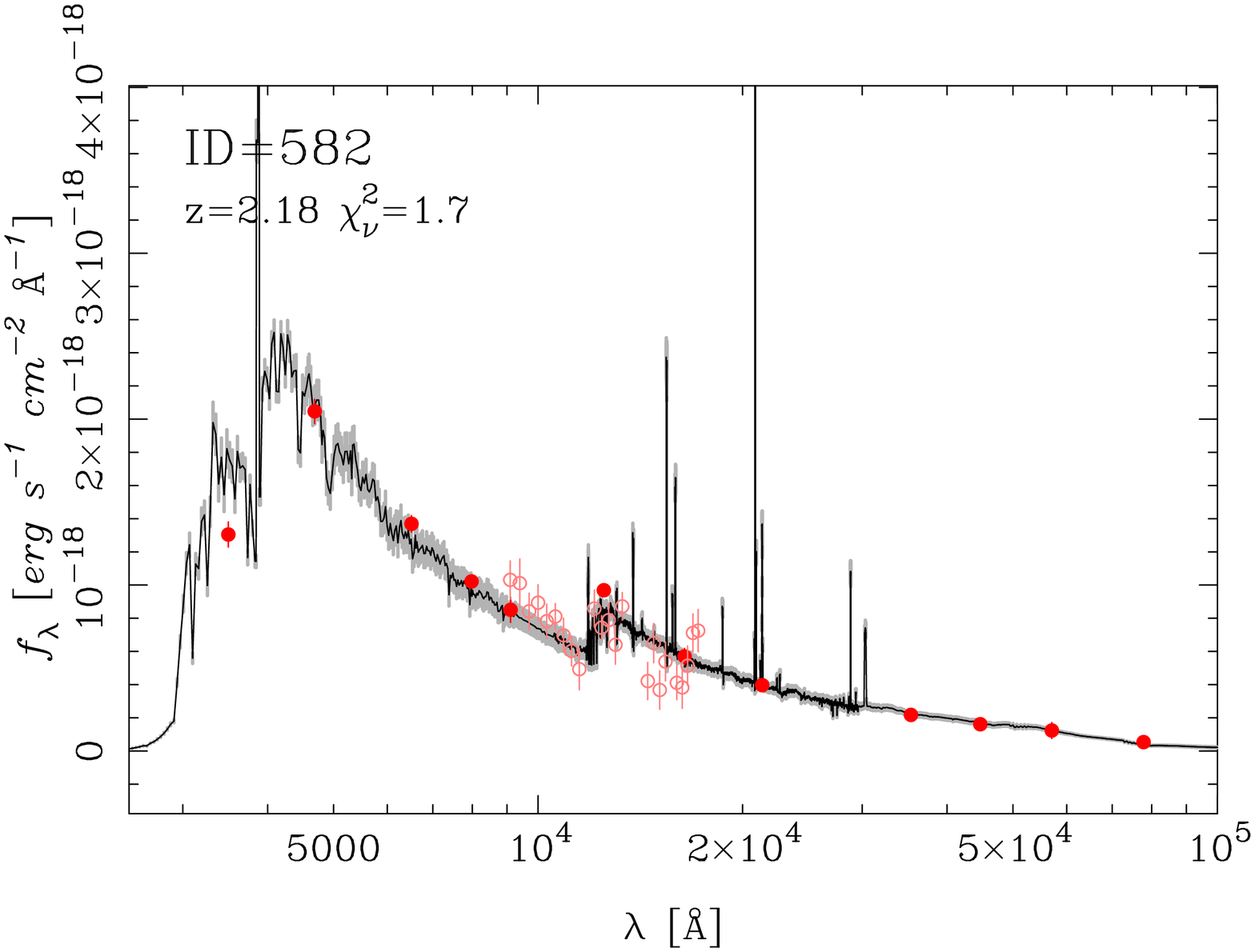}\\
\plottwo{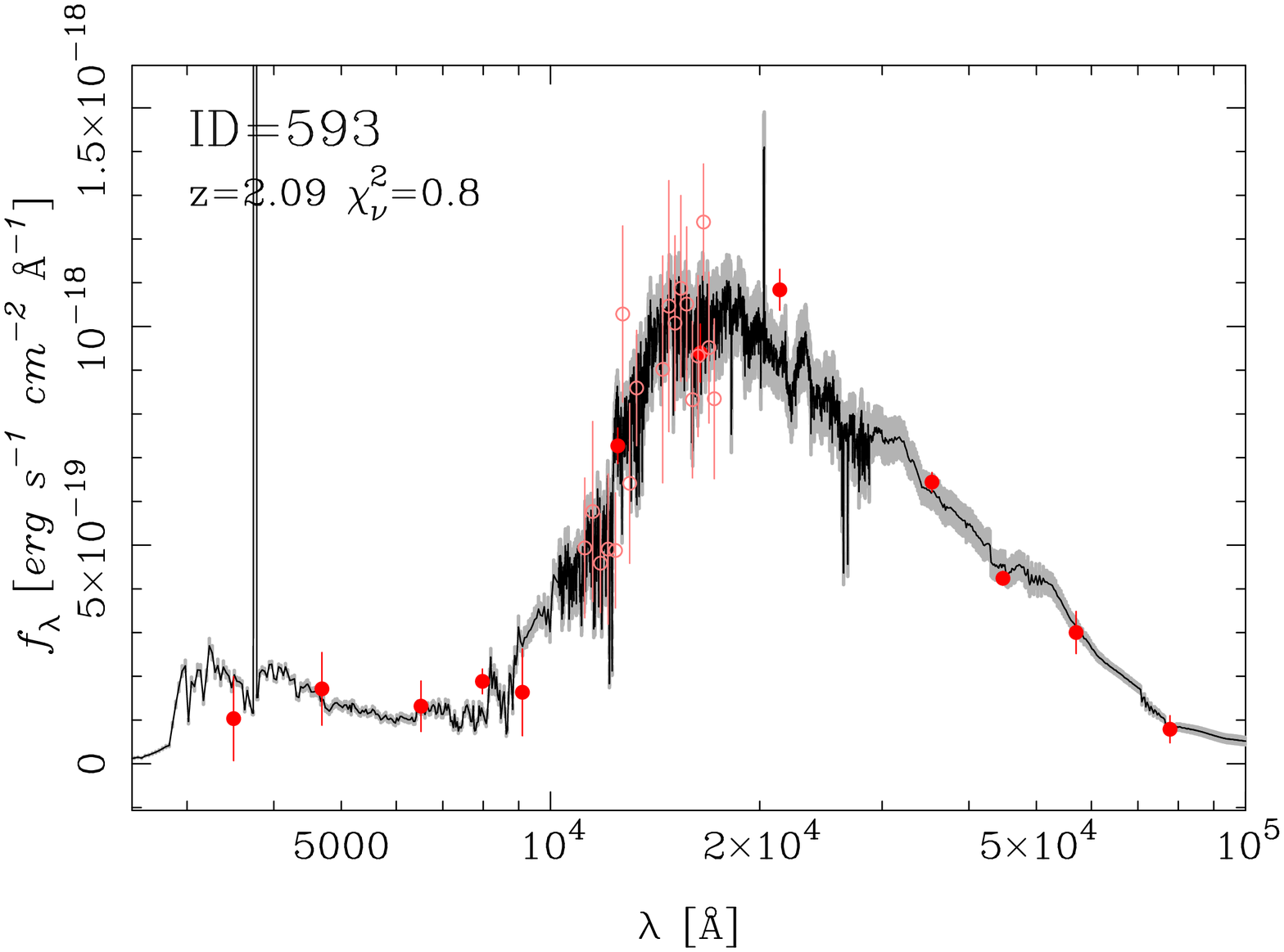}{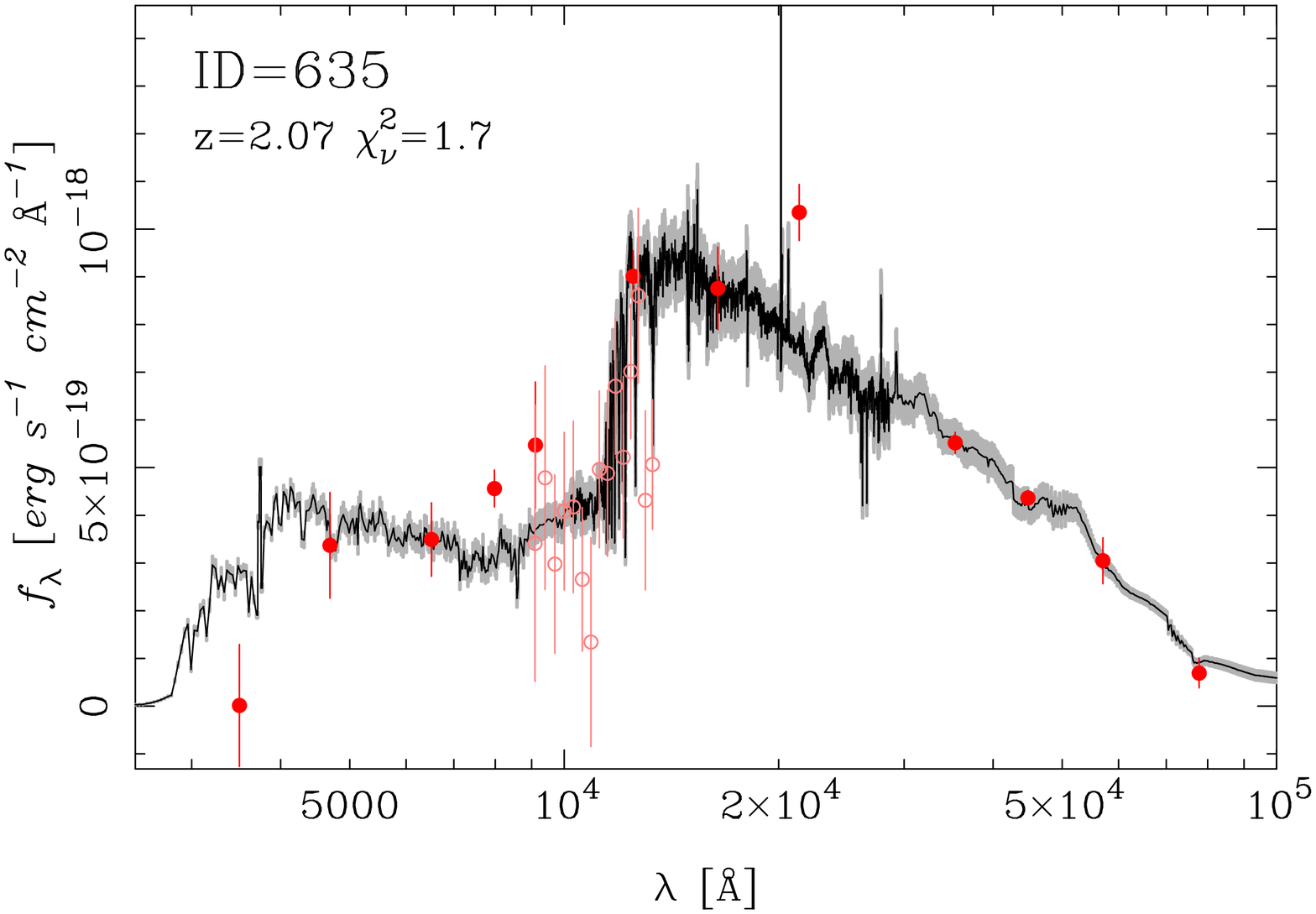}\\
\epsscale{0.5}
\plotone{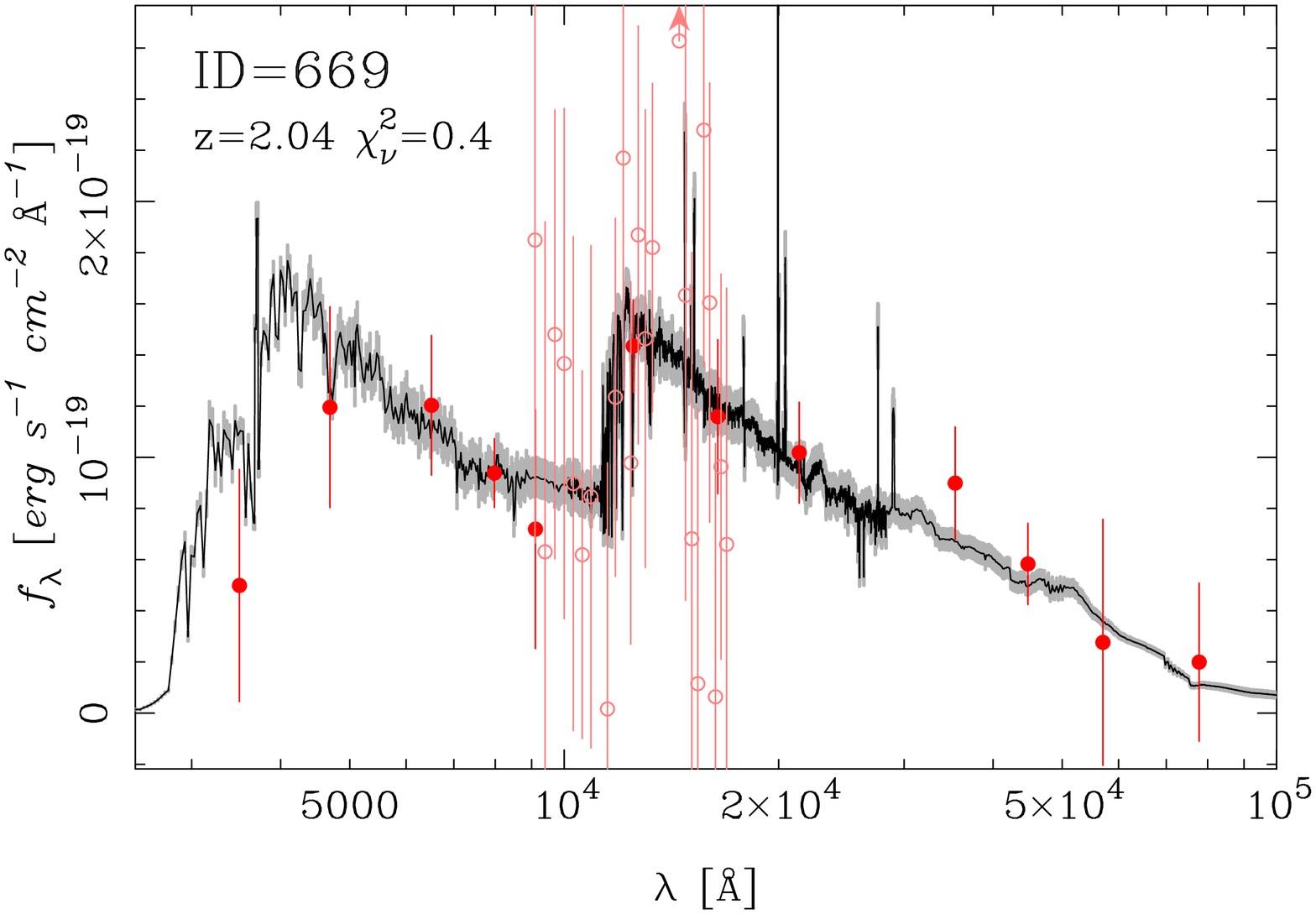}
\caption{
  As in Fig. \ref{fig:sed1}.
}
\label{fig:sed2}
\end{figure*}

We show in Figs. \ref{fig:sed1} and \ref{fig:sed2} the observed SEDs
and the best-fitting model SEDs of the member candidates.
The fits are reasonable for most objects with $\chi^2_\nu\sim1$.  
The low resolution spectra nicely trace the continuum shape and
a clear 4000$\rm\AA$ break is observed in most objects.
This continuum feature gives strong constraints on redshifts and stellar
populations of the galaxies and that is the reason why the low-S/N
near-IR spectra are very useful for $z\sim2$ galaxies (see Appendix
and also \citealt{kriek06}).

The figures illustrate the diversity of the galaxy
population in the proto-cluster; some have a blue UV slope, which
is indicative of active star formation, while some are faint
in the UV with no clear sign of active star formation.
Galaxy clusters at low redshifts are dominated by quiescent
galaxies, but this forming system  hosts both
star forming and quiescent galaxies.
We will be quantitative about this trend in what follows.

\subsection{Physical properties}

\begin{figure}
\epsscale{0.7}
\plotone{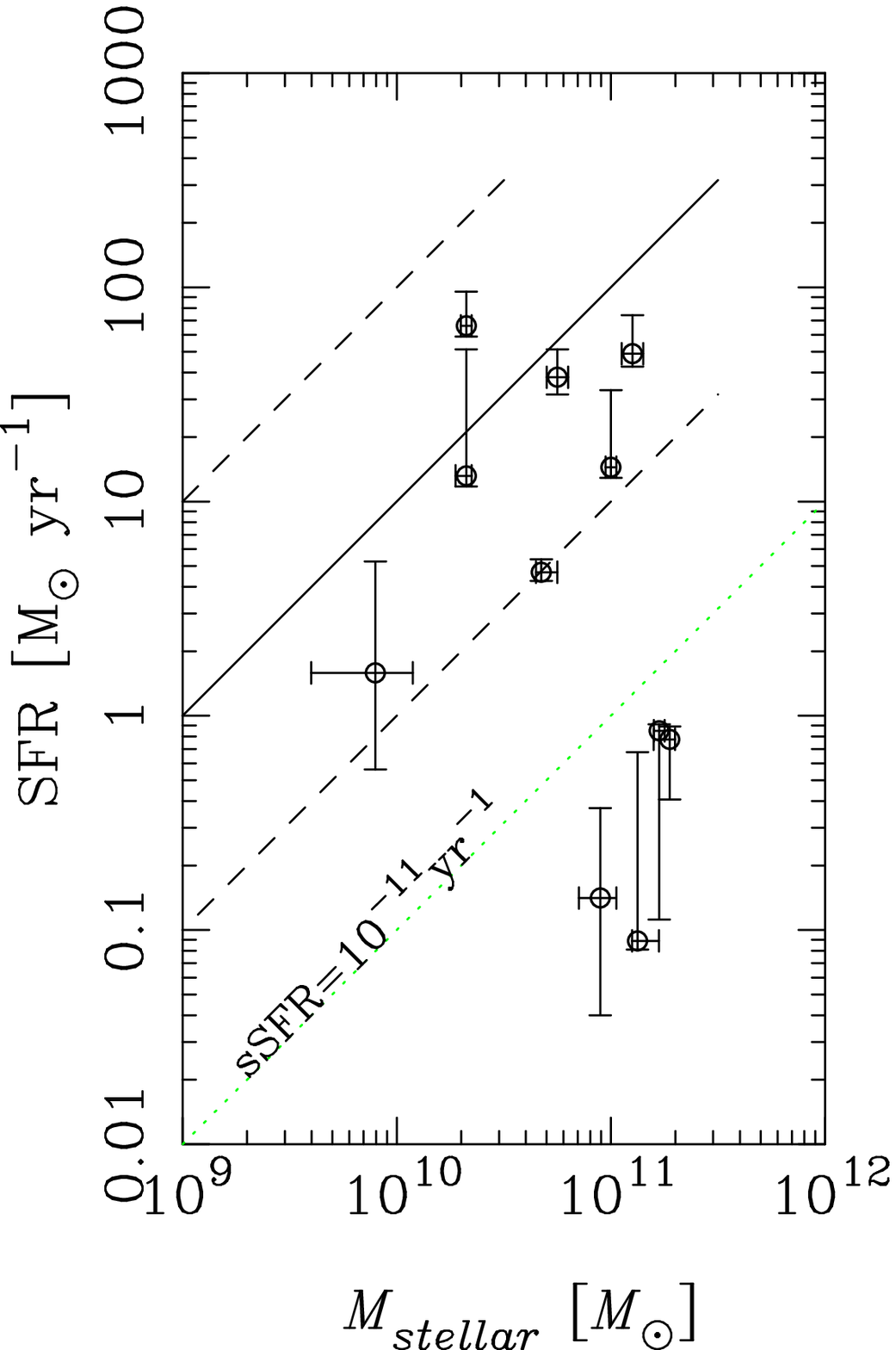}
\caption{
  SFR plotted against stellar mass.
  The solid line shows the SFR-stellar mass relation for star-forming
  galaxies at $1.5<z<2.5$ from \citet{wuyts11}.
  The dashed lines show approximate range of the relation.
  The dotted line is sSFR=$10^{-11}yr^{-1}$.
}
\label{fig:sf_sequence}
\end{figure}

We now turn our attention to the physical properties of
the member candidates.  Star formation rates (SFRs) and
stellar masses are known to show a strong positive correlation
\citep{elbaz07,noeske07} and Fig. \ref{fig:sf_sequence} shows
the relationship between the two quantities for the member candidates.
See section 3 for the details of the SED fitting and how we measure
SFR and stellar mass.  We find that the galaxies form two distinct groups:
star forming and quiescent.
Interestingly, these populations are fairly clearly separated at $sSFR=10^{-11} yr^{-1}$.
We will refer to star forming galaxies as those with $sSFR>10^{-11} yr^{-1}$
and quiescent galaxies as those with lower sSFR in what follows. 
We note that \citet{brammer09} also observed a clear separation of red and blue
galaxies in the field up to $z\sim2$.

The star formation sequence in PKS1138 might appear slightly offset towards
lower SFRs compared to the relation from the literature \citep{wuyts11}.
This might be a real trend as suggested by \citet{tanaka10}, but
it might be a systematic difference in the way SFRs are
derived from \citet{wuyts11}.
We do not attempt to pursue the issue further due to the limited statistics
(we have only 7 star forming galaxies).
If we turn to quiescent galaxies, we find that they are all massive galaxies
with $\sim10^{11}\rm M_\odot$.  There is no lower mass quiescent galaxy
in the figure, but this is likely a selection bias -- we are complete only down to
$\sim5\times10^{10}\rm M_\odot$ due to the $K_s$-band magnitude cut applied.

\begin{figure}
\epsscale{0.7}
\plotone{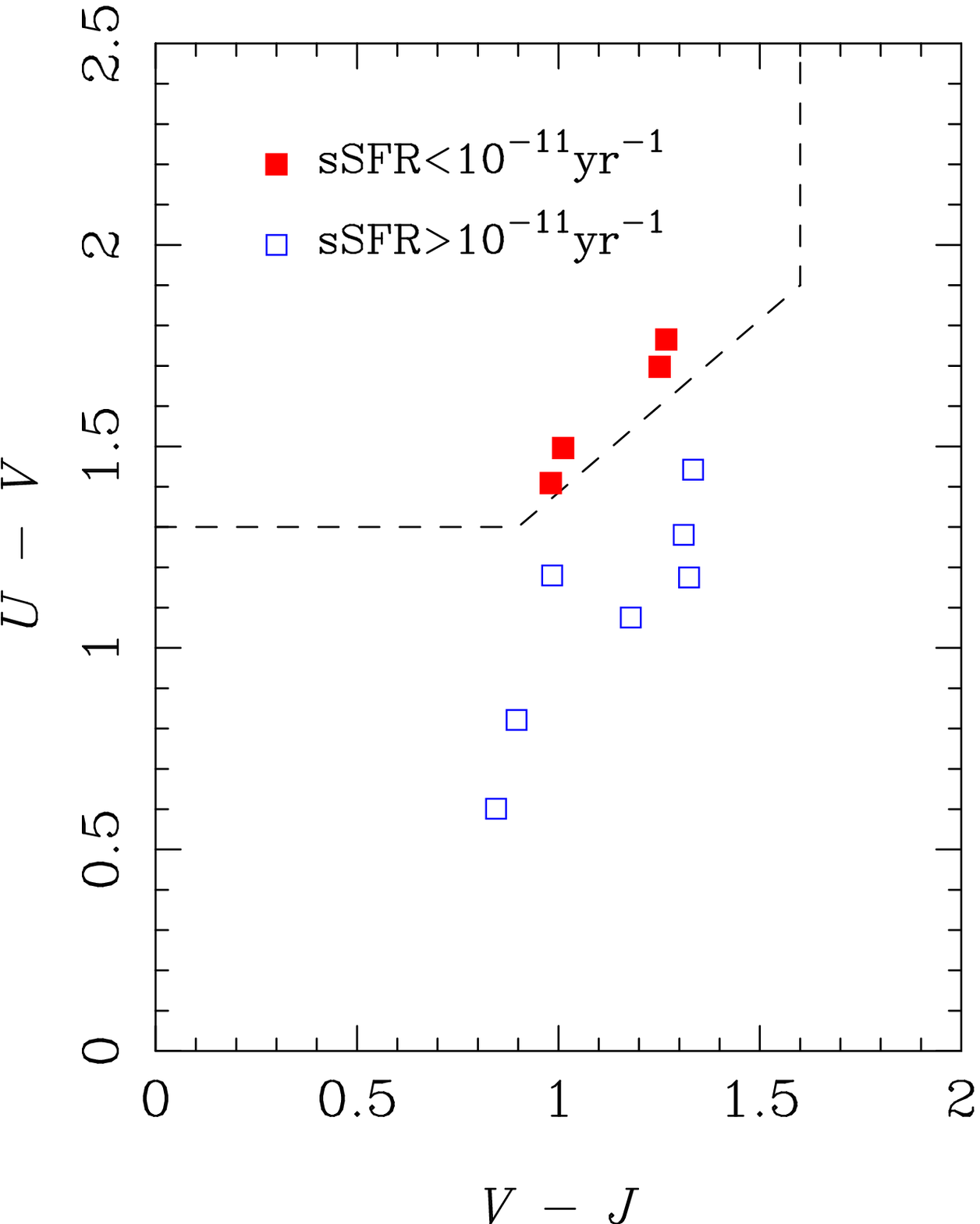}
\caption{
  Rest-frame $U-V$ plotted against $V-J$.  The dashed line is
  from \citet{williams09}, who showed that quiescent galaxies
  populate in the upper-left region.  Note that the quiescent galaxies 
  in the proto-cluster are indeed located in that region.
}
\label{fig:uv_vj}
\end{figure}

As a further check of the quiescent/star forming nature of the member galaxies,
we measure their rest-frame colors using the best-fitting models
and place them on a $U-V$ vs. $V-J$ diagram in Fig. \ref{fig:uv_vj}.
\citet{williams09} showed that the top-left corner of the diagram is
populated by quiescent galaxies and the quiescent galaxies that
we observed indeed fall in that area.
This diagram further confirms our classification of quiescent and
star forming populations based on the SED fitting.
It is tempting to estimate a fraction of
quiescent galaxies in this proto-cluster, but we choose not to do so
because our spectroscopic sampling of
the member galaxies is too small and biased (e.g., we gave
priorities to red galaxies over blue ones in the mask design).
It would be interesting to perform a less biased observation to
extend the Butcher-Oemler effect \citep{butcher84} to the epoch of
cluster formation.

\subsection{Spatial distribution}

\begin{figure}
\epsscale{1.0}
\plotone{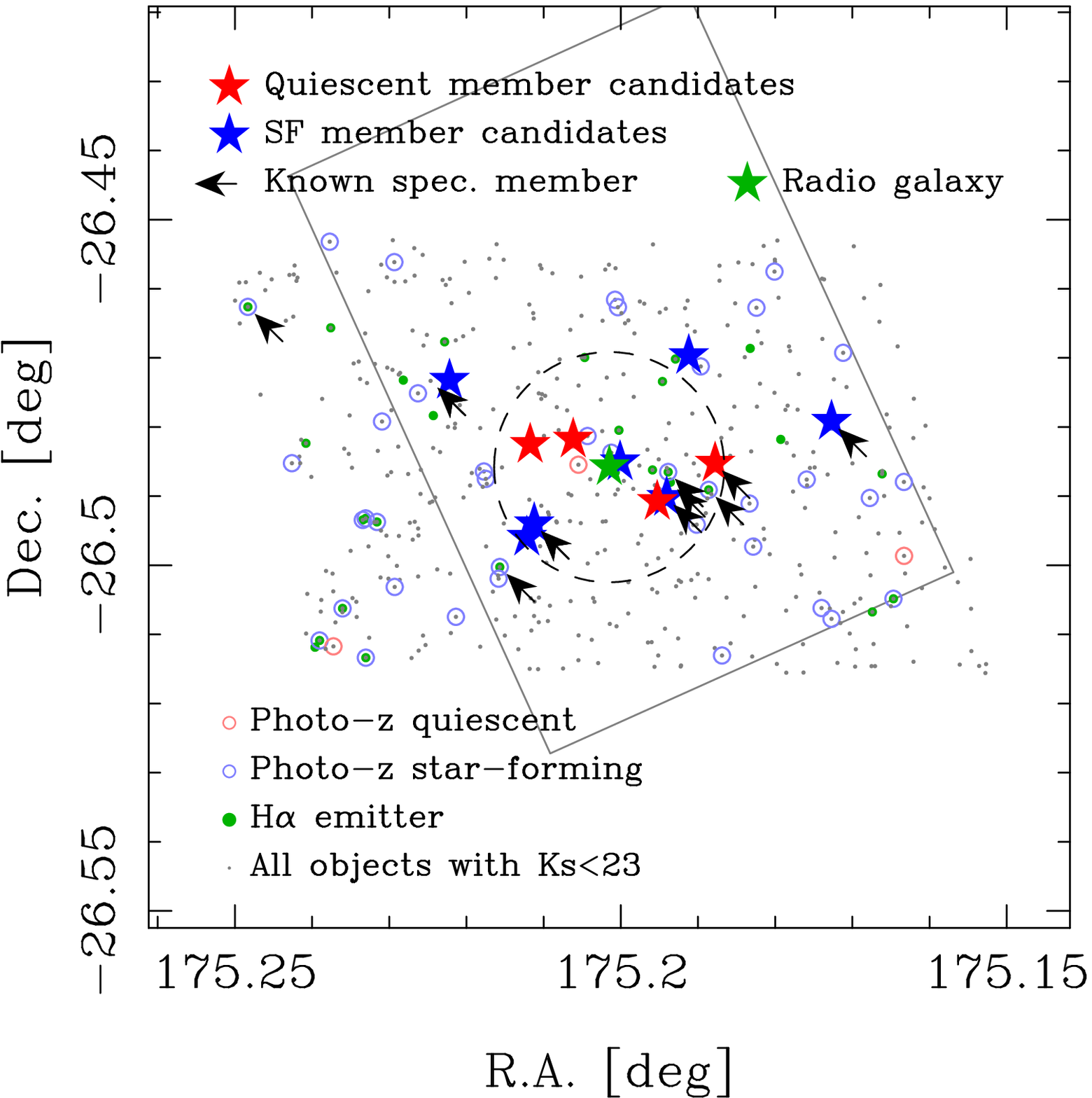}
\caption{
  Distribution of galaxies around the radio galaxy PKS1138.
  The dashed circle is 0.5 Mpc (physical) in radius.
  The rectangle shows an approximate pointing of our
  spectroscopic observation.  The pointing is optimized to
  maximize the number of high priority objects in the mask.
  The meanings of the other symbols are shown in the plot.
}
\label{fig:gal_distrib}
\end{figure}

Where are these quiescent massive galaxies located with respect to
the central radio galaxy?  Fig. \ref{fig:gal_distrib} shows the spatial
distribution of the member candidates.
We have recomputed photo-$z$'s in \citet{tanaka10} for all the objects
in the field based on the revised catalog.  The photo-$z$ objects
with $P_{cl}>0.16$ together with the H$\alpha$ emitters from \citet{koyama13}
are shown in the figure for reference.
We find that the spectroscopically observed quiescent members
shown as the red stars seem to be clustered
around the radio galaxy and they are all located within 0.5 Mpc
from the radio galaxy.  This is unlikely due to 
a selection bias introduced in the mask design because 
we did not prioritize red galaxies in the center over those
in the outer parts.  Although the spatial area we explore 
is not large, this plot might indicate that massive quiescent
galaxies already dominate the central part of a forming cluster.

As mentioned above, we suffer from near-foreground/background
contamination due to the limited accuracy of $z_{specphot}$.
Here, we argue that the concentration of the quiescent galaxies
is not due to the contamination.
\citet{brammer11} estimated a number density of quiescent galaxies
with stellar mass of $>10^{11}\rm M_\odot$ to be $\sim(5\pm2)\times10^{-5}\rm\ Mpc^{-3}$
at $z\sim2.1$.
Assuming that we sample galaxies at $2.0<z<2.3$ with Eq. 1,
we expect $\sim0.3\pm0.1$ massive quiescent galaxies within the circular
aperture shown in  Fig. \ref{fig:gal_distrib}, instead we observe four.
Even over the entire field shown in Fig. \ref{fig:gal_distrib},
we expect $\sim1.3\pm0.5$ such galaxies.
Of course, these numbers are 
subject to cosmic variance, but it is in any case unlikely that
we significantly suffer from the contamination and the concentration
of the massive quiescent galaxies around the central radio galaxy
is likely real.
We will focus on these massive quiescent galaxies in Section \ref{sec:stacked_quiescent}
assuming that the contamination is negligible.
As for star forming galaxies, 6 out of 7 galaxies are H$\alpha$
emitters (Section \ref{sec:ha_emitters}) and thus the contamination
of field star forming galaxies is also likely small.

\subsection{Color-magnitude diagram}

\begin{figure}
\epsscale{1.0}
\plotone{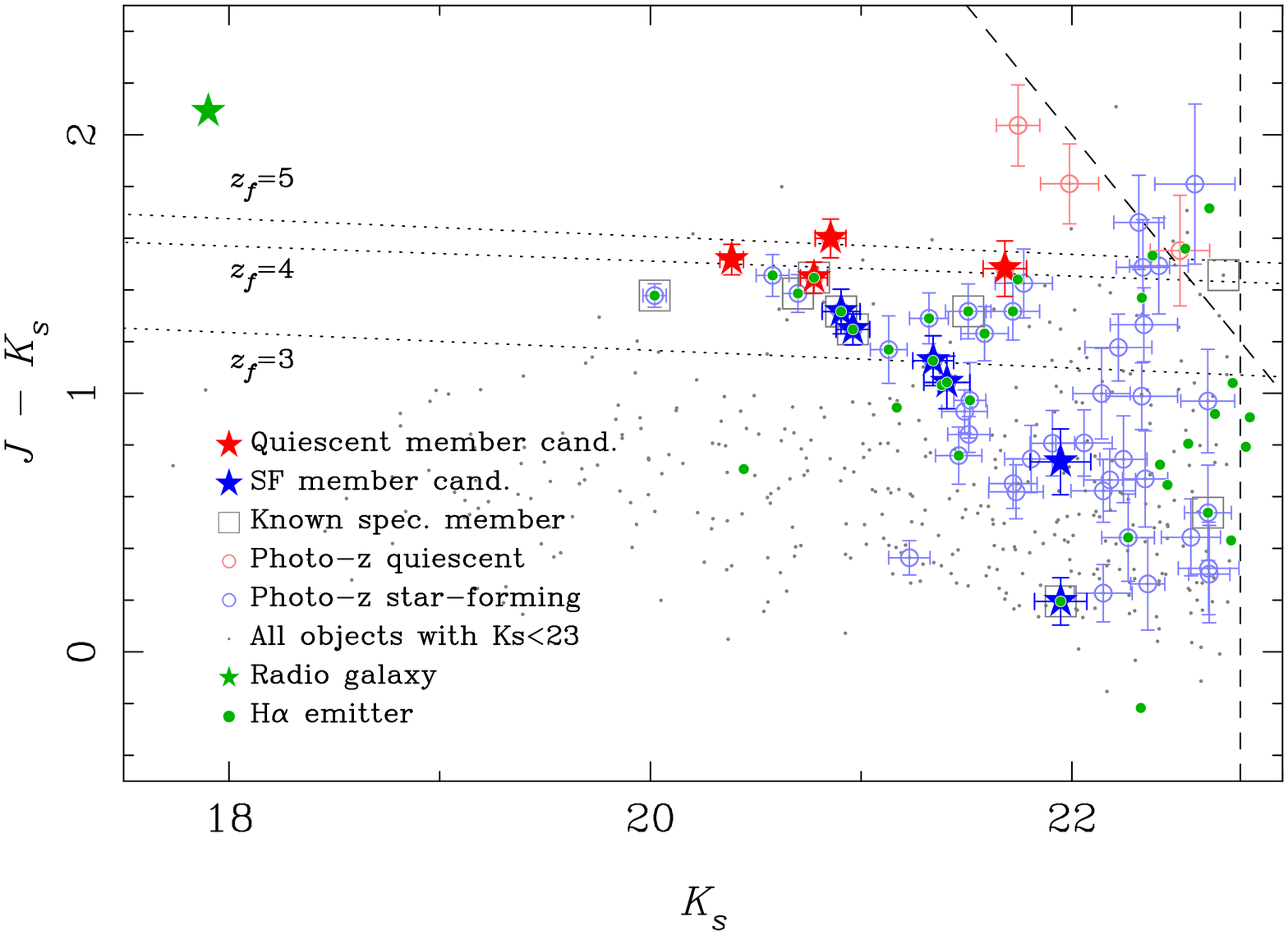}
\caption{
  $J-K_s$ plotted against $K_s$.  The vertical and slanted
  dashed lines show $K_s=23$ ($\sim5\sigma$ limit) and
  $5\sigma$ limiting color, respectively.  The dotted lines are the model
  red sequence formed at $z_f=3$, 4, and 5 \citep{tanaka10}
  as indicated in the plot.  The meanings of the other symbols
  are the same as in Fig. \ref{fig:gal_distrib} and are shown
  in the plot.
}
\label{fig:cmd}
\end{figure}

Galaxies in low-redshift clusters are known to form a tight
red sequence on a color-magnitude diagram (e.g., \citealt{bower92}).
It would be interesting to examine whether the quiescent galaxies form
a red sequence in a progenitor system such as the one studied here.
We present a color-magnitude diagram in Fig. \ref{fig:cmd}.

Most of the bright galaxies have a relatively red color,
but the observed red sequence is not very tight and is
contaminated by star forming galaxies as shown by \citet{koyama13}.
This is perhaps what one would expect to observe in a forming cluster.
Interestingly, however, the reddest part of the sequence is
populated by the quiescent galaxies with sSFR$<10^{-11}\rm\ yr^{-1}$,
forming a weak red sequence.   
We note that there are a few photo-$z$ selected, quiescent galaxies
with very red colors with $K_s\sim22$, but they have large photometric
errors and their sSFRs are actually consistent with
star forming galaxies within $1\sigma$.  Only the member candidates
selected with $z_{specphot}$ shown as the red stars are quiescent
at a high significance.
This is the first spectroscopic
confirmation of red sequence formed by quiescent galaxies
in a proto-cluster.
This observation adds a support to the claim that
the red sequence is being formed in this system \citep{zirm08}.
We may expect the red star forming galaxies will eventually
stop forming stars and form a more prominent red sequence at a later epoch.
If we perform a linear fit to the 4 quiescent galaxies by fixing the
slope to that of the model red sequence at $z_f=4$ (i.e., only the offset
is allowed to float), we obtain $z_f=4.1\pm0.2$.  We will come back
to this number in the discussion section.

\vspace{0.3cm}

To sum up, we have performed a detailed analysis of the SEDs 
and physical properties of the member candidates of
the proto-cluster using the broad-band photometry and low resolution
spectra simultaneously.  We find that quiescent galaxies and
star forming galaxies co-exist in this proto-cluster and
the quiescent galaxies seem to form a weak red sequence.
The quiescent galaxies are spatially concentrated around
the central radio galaxy.  This might suggest that the red sequence
first appears when a system is collapsing to form a gravitationally
bound system.  
We examine stellar populations of these galaxies in further detail
using the MOIRCS spectra in the following sections.

\section{Stacked spectrum of star forming members}
\label{sec:stacked_sf}

Rest-frame optical spectra can place tight constraints on the stellar
populations of galaxies through an absorption/emission line analysis.
As mentioned earlier, the typical S/N of our individual MOIRCS spectra
is not good enough for such an analysis.  We here attempt to stack
the spectra to obtain a higher S/N spectrum.
Although the primary focus of the paper is on quiescent galaxies (Section 6),
we briefly discuss the nature of star forming galaxies for completeness
in this section.

Spectroscopic redshifts are available for 2 of the 7 star forming
galaxies in the proto-cluster and we assume that the rest are all
located at the cluster redshift.  One object is excluded from this
analysis because the object is an X-ray source (ID=399) and we do
not want to bias the emission line anlalysis below.
All the spectra of the remaining 6 star forming galaxies are
shifted to rest-frame wavelength, normalized over the entire wavelength,
and combined with inverse-variance weights in a rest-frame wavelength
bin of $3.6\rm\AA$, which is a half of $\Delta \lambda/\lambda$ of
our observation.

\begin{figure*}
\epsscale{1.0}
\plotone{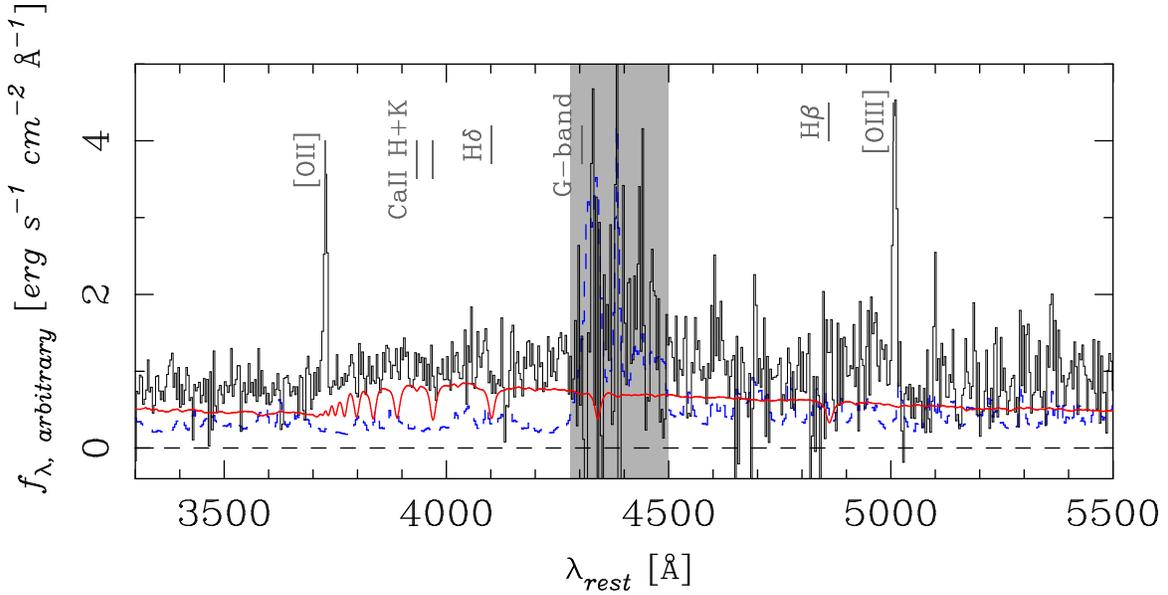}
\caption{
  Stacked spectrum of the 6 star forming galaxies in the cluster.
  The dashed spectrum is a noise spectrum and the solid, smooth
  spectrum is the best-fit model spectrum shifted downwards
  for clarity (see text for details).  Some of the prominent
  spectral features within the probed wavelength range are indicated.
  The shaded area is strongly affected by the atmospheric absorption.
}
\label{fig:combspec_sf}
\end{figure*}

We present the stacked spectrum of 6 star forming galaxies in
Fig. \ref{fig:combspec_sf}.
We observe prominent emission lines ({\sc [oii]} and {\sc [oiii]})
on top of a relatively blue continuum with a weak Balmer break.
We fit the whole spectrum with the updated \citet{bruzual03} model
templates assuming the Chabrier IMF and the exponentially declining
star formation histories as above. Here, we allow age,
extinction ($\tau_V$), and star formation time scale to vary
and the best-fitting model spectrum is shown in Fig. \ref{fig:combspec_sf}.
The derived physical parameters are: age=$0.16^{+0.10}_{-0.01}$ Gyr,
$\tau_V=1.20^{+0.05}_{-0.20}$, and $\tau=0.10^{+0.35}_{-0.00}$ Gyr.
This young age indicates active star formation in these galaxies.

\begin{figure*}
\epsscale{1.0}
\plotone{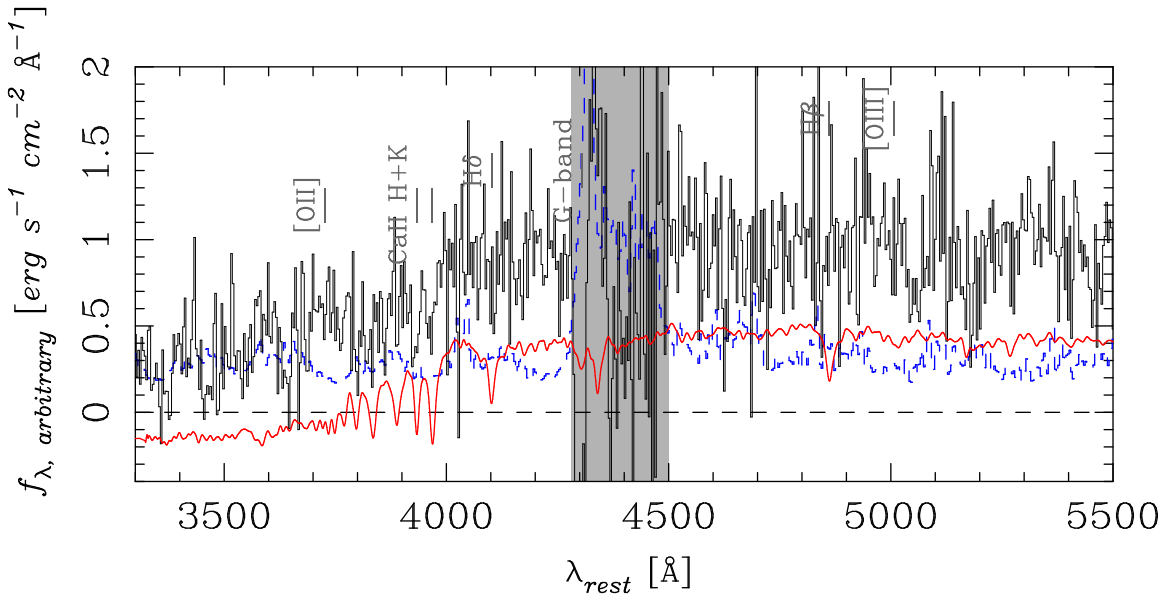}
\caption{
  As in Fig. \ref{fig:combspec_sf}, but for the stacked spectrum
  of the 4 quiescent galaxies in the cluster.
}
\label{fig:combspec}
\end{figure*}

We subtract the best-fit model spectra to measure emission line fluxes.
H$\beta$ emission is not clearly seen, but if we assume that the spike
at the wavelength consistent with H$\beta$ is a real feature, we measure
{\sc [oiii]$\lambda5007$}/H$\beta=2.9\pm0.4$, a large ratio not commonly
observed in massive star forming galaxies at $z=0$.  This indicates
either the galaxies are relatively metal-poor galaxies or they host AGNs.
It is hard to discriminate these two possibilities without the {\sc [nii]}/H$\alpha$
ratio, but the latter may be the case as discussed below.

We use the $R_{23}$
method to estimate their gas-phase metallicity (see \citealt{kewley08} and
references therein).  The $R_{23}$ is double-valued with metallicity and
we cannot find a unique solution, but we measure
$\rm 12+\log[O/H]=8.70^{+0.09}_{-0.12}$ for the upper branch and 
$8.24^{+0.10}_{-0.08}$ for the lower branch using the calibration by \citet{kobulnicky04}.
The star forming galaxies studied here typically have stellar mass of
$\sim6\times10^{10}\rm\ M_\odot$.  \citet{erb06} reported that star forming
galaxies at $z\sim2$ of similar stellar mass have $\rm 12+\log[O/H]=8.52^{+0.06}_{-0.05}$.
Their metallicity is based on the \citet{pettini04} calibration and if we translate
it into the \citet{kobulnicky04} calibration following \citet{kewley08},
we obtain $\rm 12+\log[O/H]=8.92\pm0.05$ plus an uncertainty of 0.067 dex
from the conversion.
Our estimate even for the upper branch is lower than this by
$\Delta\log\rm[O/H]\sim-0.2$.  This might be a real trend, but we would
rather interpret it as a hint of AGN contamination.
Although we have excluded an X-ray source from the stacking, some of the other
objects may host AGNs, which tend to lower the metallicity estimates from $R_{23}$.
A caveat to the analysis here is that we use the extinction from
the spectral fit to correct for the dust extinction on the emission lines,
which may not be a very accurate correction.
The  {\sc [nii]}/H$\alpha$ ratio is needed to draw a firm conclusion
on the prevalence of AGNs among star forming galaxies in this proto-cluster.

\section{Stacked spectrum of quiescent members}
\label{sec:stacked_quiescent}

We now turn our attention to quiescent galaxies.  Given the likely
forming red sequence, it is interesting to study their stellar populations
and formation histories in detail and that will be the focus of the rest of the paper.
We follow the same procedure for the stacking as described above,
except that we normalize the spectra at $\rm 4500\AA<\lambda_{rest}<5500\AA$
(i.e., above the break), where we have better S/N than below the break.
We perform inverse-variance weighted stacking as done in the previous
section.  This weights towards
brighter galaxies, but we have confirmed that our
argument below does not change if we perform straight-mean stacking.

We assume that all the quiescent galaxies are located at the radio
galaxy redshift of $z=2.16$.  We slightly tweak 2 of the 4 redshifts
($\delta z\sim0.01$) in which a tentative hint of absorption features is
observed in the individual spectra. 
We present the stacked spectrum in Fig. \ref{fig:combspec}.
The continuum is fairly flat at $>4000\rm\AA$, while the red
continuum is seen at shorter wavelengths and the 4000\AA\ break
is clearly observed.  Interestingly, there
is a hint of CaII H+K absorption in the spectrum.  This is not
a convincing detection, but if real, this would be the first
detection of the feature in quiescent galaxies in a proto-cluster
at a high redshift.  Strong emission lines are absent in the spectrum,
which is in stark contrast to the star forming galaxies in Fig.
\ref{fig:combspec_sf}, confirming the quiescent nature of the galaxies.

\begin{figure}
\epsscale{1.0}
\plotone{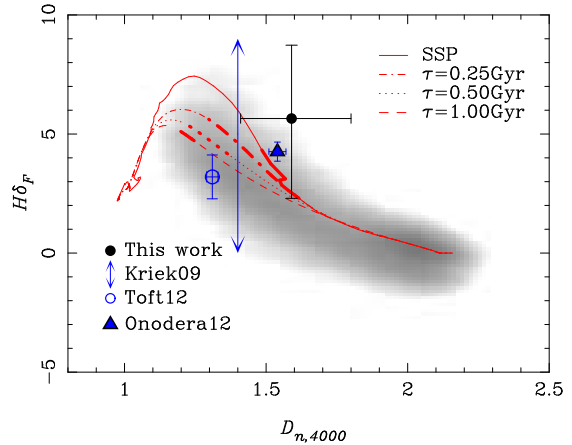}
\caption{
  H$\delta_F$ plotted against $D_{n,4000}$. The gray scale shows
  $z=0$ galaxies drawn from SDSS.  The large point is the quiescent
  galaxies in PKS1138 and the arrow indicates $D_{n,4000}$ of
  a massive quiescent galaxy in the field at $z=2.2$ \citep{kriek09}.
  Another massive field galaxy at $z=2$ from \citet{toft12} is shown
  as the open circle.    The triangle is a stacked spectrum of
  field galaxies located at lower redshifts
  ($z\sim1.6$; \citealt{onodera12}).  The model
  tracks are shown as the curves.  The thick parts of the curves
  show the range of $3<z_{f}<5$ observed at $z=2.16$.
}
\label{fig:hd_d4000}
\end{figure}

We make a further attempt to constrain the stellar population
of the quiescent galaxies by measuring the strengths
of the 4000$\rm\AA$ break and the H$\delta$ absorption of the stacked
spectrum.  For these indices, we adopt the definition of $D_{n,4000}$
by \citet{balogh99} and H$\delta_F$ by \citet{worthey97}, respectively.
We measure $D_{n,4000}=1.59_{-0.18}^{+0.21}$ and H$\delta_F=5.65_{-3.35}^{+3.08}$
from the stacked spectrum.
Due to the large noise, these indices cannot be precisely
measured, but they still give interesting constraints on the formation
of the cluster massive galaxies as discussed below.
We compare this stacked spectrum with (a) quiescent galaxies in the field
at similar redshifts and (b) stellar population synthesis models.
We also show the distribution of the galaxies at $z=0$ drawn from
the Sloan Digital Sky Survey (SDSS; \citealt{york00}) for reference.
We show a  H$\delta_F$ vs. $D_{n,4000}$ diagram in Fig. \ref{fig:hd_d4000}
and make the two comparisons here.

\subsection{(a) Quiescent galaxies in the field}

\citet{onodera12} studied color-selected massive galaxies in the COSMOS field
using the same instrument as ours.  The redshifts of their galaxies spread over
$1.4\lesssim z \lesssim 1.8$ with masses around $\sim10^{11}\rm M_\odot$.
Although their objects are located at lower redshifts than ours, 
we plot the $D_{n,4000}$ and H$\delta_F$ indices of their stacked spectrum in
Fig. \ref{fig:hd_d4000} as the filled triangle.  
Their spectral indices have very small statistical uncertainties 
due to the high S/N of the spectrum.
Our stacked object has a rather large uncertainty and its H$\delta_F$ and
$D_{n,4000}$ is consistent with \citet{onodera12}, but let us repeat that
their objects are located at lower redshifts.

\citet{kriek09} performed extremely deep near-IR spectroscopy of a massive,
quiescent galaxy at $z=2.2$ with a stellar mass of $2\times10^{11}\rm M_\odot$.
They measured $D_{n,4000}=1.40^{+0.03}_{-0.03}$, which is bluer than
our quiescent objects at $1\sigma$.
\citet{kriek09} did not quote H$\delta_F$ and hence further
comparisons cannot be made.  Another field galaxy is drawn from \citet{toft12},
who carried out an X-shooter observation of a massive galaxy at $z=2$ with
a stellar mass of $\sim2\times10^{11}\rm\ M_\odot$.  The galaxy is also bluer
compared to the quiescent proto-cluster galaxies.

Although the sample is very limited at this point, the two
quiescent galaxies in the field
at $z\sim2$ seem to be bluer than those in the proto-cluster.
The bluer colors of  field quiescent galaxies is 
expected in the framework of the hierarchical universe.
The galaxy formation occurs in a biased fashion and we expect
galaxies in clusters start to form earlier than field galaxies,
and hence the proto-cluster galaxies appear older (redder) than
the field galaxies at similar epoch.
Unfortunately, however, the current statistics are too poor to draw
any firm conclusion
on the environmental variation in the spectral indices.

\subsection{(b) Stellar population synthesis models}

The redshift of $z=2.16$ corresponds to 3 Gyr since the Big Bang.  The location of
the red sequence in lower redshift clusters indicates the 
formation redshift of  $3\lesssim z_f\lesssim5$ (e.g., \citealt{stanford98}),
which is only 1-2 Gyr prior to $z=2.16$.  This fact allows us to put
a tight constraint on the formation time scale of cluster massive
galaxies.

We use an updated version of \citet{bruzual03} population synthesis
code with improved treatment of the thermally pulsating AGB stars and
assume the Chabrier IMF as in our spectrophotometric fits.
Because the observed galaxies are massive galaxies with stellar
mass of $\sim10^{11}\rm M_\odot$, it is reasonable to assume solar
metallicity.
In fact, \citet{toft12} found that a similarly massive galaxy
at $z=2$ has metallicity consistent with solar.
We do not consider dust extinction here because the spectral indices
we measure are relatively independent of extinction.
The only free parameters considered here are the exponential decay time
scale ($\tau$) for the star formation history and age, which is time
since the onset of star formation.  All the model spectra are
smoothed to the instrumental resolution of our observation.

We show in Fig. \ref{fig:hd_d4000}  model tracks for $\tau=0,\ 0.25,\ 0.5,$
and 1.0 Gyr.  The thick parts of the curves indicate the formation redshift
range of $3<z_f<5$ observed at $z=2.16$, and thus the stacked spectrum
should be compared with these thick parts.
The quiescent galaxies in the proto-cluster at $z\sim2.2$ can be reproduced
by the models with $\tau=0$ and 0.25 Gyr.  The models with $\tau=0.5$ and 1 Gyr
are inconsistent at 1 and $2\sigma$ levels, respectively.  This suggests that,
under the assumption of the $\tau$-model, the star formation time scale
of massive quiescent galaxies in the proto-cluster has to be short, $\lesssim0.5$ Gyr.
If we allow a higher formation redshift of $z_f=10$, $\tau=0.5$ Gyr is consistent
with the observation, while $\tau=1$ Gyr is still ruled out at $1.5\sigma$.
The extra time we gain by allowing higher formation redshifts is short
and the short star formation time scale inferred here is not strongly
dependent on the choice of the maximum $z_f$.

While this spectral analysis is robust against flux calibration uncertainties
(the indices are measured in narrow wavelength windows), it provides
a limited constraint on $\tau$ and $z_f$.  This is primarily because our
H$\delta_F$ is essentially unconstrained and $z_f$ and $\tau$ are degenerate.
We perform a further analysis to constrain these parameters
using the full stacked spectrum. 
We fit the whole spectrum in the same manner as described in Section \ref{sec:stacked_sf}.
The derived physical parameters are: age=$1.02^{+0.13}_{-0.02}$ Gyr,
$\tau_V=0.40^{+0.10}_{-0.22}$, and $\tau=0.10^{+0.02}_{-0.01}$ Gyr.
The quiescent galaxies are older and have less dust compared to
the star forming galaxies (Section \ref{sec:stacked_sf}),
clearly illustrating more evolved stellar populations.
The short $\tau$ is in agreement with the constraint
from Fig. \ref{fig:hd_d4000} ($\tau\lesssim0.5$ Gyr) and
the inferred age translates into  $z_f=3.12^{+0.18}_{-0.02}$.

The derived parameters have small formal errors on age (i.e., $z_f$)
and $\tau$, but
the spectral fit here is subject to systematics such as flux calibration
uncertainty.  We try to gauge the systematic uncertainties in our
results here by comparing with the formation redshift inferred from
the location of the red sequence in Fig. \ref{fig:cmd}.
We recall that we have measured $z_f=4.1\pm0.2$ from the red sequence,
which is higher than that from the spectral fit.
But, the difference is in part due to the dust extinction applied here.
The amount of extinction is not too high, but if we perform
the same spectral fit by fixing extinction to 0, we obtain
age=$1.28^{+0.02}_{-0.14}$ Gyr and $\tau=0.10^{+0.01}_{-0.06}$ Gyr.
This age translates into $z_f=3.50^{+0.05}_{-0.21}$.
The age and dust extinction are known to degenerate and the latter fit
with $\tau_V=0$ is likely more robust.
The formation redshift  remains essentially the same if
we further assume $\tau=0$ as in the red sequence analysis
(note that best-fit $\tau$ quoted above is fairly small).
This $z_f$ is still inconsistent with $z_f=4.1$ from the red sequence
and we deem that the redshift difference of $\Delta z_f=0.6$
($\sim0.3$ Gyr)
is a level of systematics in our analysis here.
The flux calibration uncertainty in the spectrum and
the zero-point calibration uncertainty in the broad-band photometry
will be the main sources of systematics.
It would be fair to say that the formation redshift is
$3\lesssim z_f\lesssim4$.
We can apply the same argument to the formation time scale;
$\tau$ from the full spectral fitting is smaller than that
from $D_{n,4000}$, and we use that result to gauge the level
of systematics.
To be conservative, we choose to simply place an upper limit;
$\tau\lesssim0.5$ Gyr.

We have considered systematics in observations, but there is another
important source of systematics, which is the stellar
population synthesis model.  All the models used here are based on
the updated \citet{bruzual03} code.  There are a number of uncertainties
in stellar population synthesis models
as summarized in \citet{conroy09}.  We do not reiterate
them here, but a possibly important role of thermally pulsating AGB
stars claimed by \citet{maraston05} is worth mentioning because
these stars contribute significantly to the near-IR SED at an age of
$\sim1$ Gyr, which is about the estimated age of the quiescent galaxies.
Here, we argue that our results are robust against this uncertainty.
Our spectral analysis is performed in
the rest-frame optical wavelengths, where these unstable stars do
not significantly contribute.  In fact, the evolution of $D_{4000}$
around 1 Gyr is very similar between the \citet{bruzual03} model
and \citet{maraston05} model.  Furthermore, a prominent role of thermally pulsating
AGB stars has been questioned by several authors (e.g.,
\citealt{conroy10,kriek10,zibetti13}).  For these reasons, we do not consider
the thermally pulsating AGB stars are a serious concern in our analysis.
Of course, the evolution of $D_{4000}$ as well as overall spectral shape
itself is subject to systematics \citep{conroy10}
and it remains one of the major uncertainties at this point.
Attenuation curve is not
a major concern here because we specifically focused on quiescent
galaxies.  The initial mass function is assumed to be \citet{chabrier03},
but if we assume \citet{salpeter55}, stellar mass and SFR increase
by roughly a factor of 2 and the other parameters do not significantly
change.  
The assumption of solar metallicity may also be a concern.
\citet{toft12} estimated metallicity consistent with
solar for a massive field galaxy at $z=2$ albeit with a large uncertainty.
Direct, precise metallicity measurements of our proto-cluster galaxies
will be essential, but we note that, even if we assume (somewhat unrealistic)
super-solar metallicity with $Z=0.05$, a short formation time scale ($\tau\lesssim1$ Gyr)
is still required to reproduce the observed red spectrum.
For sub-solar metallicity, the time scale becomes even shorter.

We have summarized a number of sources of systematics above and
we argue that our results do not significantly suffer from most of them.
The only major concern is the uncertainty in the overall spectral evolution of
the models in the optical at ages of a few Gyr, which is difficult to
quantify at this point.  We should bear it in mind, but 
it is unlikely that the inferred short formation time scale changes
by as much as $\sim1$ Gyr, which is needed to bring the observation
consistent with recent simulations as discussed below.
We conclude that, under a number of assumptions, the quiescent
galaxies in PKS1138 form at $3\lesssim z_f\lesssim4$ with a formation time scale
of $\lesssim0.5$ Gyr.
The remaining question is, how can these galaxies grow to
$\sim10^{11}\rm M_\odot$ on such a short time scale?

\section{Summary and Discussion}

We have performed deep near-IR spectroscopy of a z=2.16 proto-cluster.
We find that quiescent galaxies with masses $\gtrsim10^{11}\rm\ M_\odot$
already appear in the forming cluster.
We spectroscopically confirm the red sequence of quiescent galaxies
in a proto-cluster for the first time.
But, the observed red sequence is weak and the presence of red star
forming galaxies suggests that the red sequence is being formed.
The stacked spectrum of the star forming galaxies exhibits prominent
emission lines on top of the blue continuum, indicating active star formation.
However, their inferred gas-phase metallicity is low compared to that of
field galaxies of similar mass at similar redshift, which we interpret
as an indication of AGN activities in the star forming galaxies.
On the other hand, the stacked spectrum of the quiescent galaxies
shows a clear 4000\AA\ break with a hint of a CaII H+K feature,
indicating an evolved stellar population.  Detailed spectral analyses
based on the stellar population synthesis models suggest that
the quiescent galaxies form at $3\lesssim z_f\lesssim4$ with a formation time
scale of $\lesssim0.5$ Gyr.

This short formation time scale is consistent with what has been
indicated by stellar absorption studies of massive ellipticals
in the local universe (e.g., \citealt{thomas99,thomas05});
the observed $\alpha$ element enhancement suggests that their major
episode of star formation has to be short so that type Ia supernovae
do not significantly contribute to the overall metal enrichment.
The exact delay time of type Ia supernovae is still uncertain, but
Eq. 4 of \citet{thomas05} gives a star formation time scale
of 0.4 Gyr for a $10^{11}\rm M_\odot$ galaxy, which is consistent
with our finding.  One of the main criticisms of this interpretation
is that it is possible to attribute the $\alpha$-enhancement
to an IMF variation.
But, our observation at $z\sim2$ presented in this paper
provides independent evidence for the rapid formation.

\citet{naab07} performed SPH simulations of massive galaxy formation
in the cosmological context.  They show that most star formation happens
at high redshifts, while a significant mass assembly happens at later
times, $z<1$.  This nicely reproduces the observation that the stellar
populations of massive early-type galaxies today are old.  However, the overall
star formation time scale of the massive galaxies ($\sim10^{11}M_\odot$)
in \citet{naab07} is $\tau\sim1.5$ Gyr,
which is longer than the time scale inferred from our observations.
\citet{johansson12} extended the work by including stellar feedback,
but the star formation time scale seems to remain similar (see their Fig. 2).
This might indicate that there is a missing physical process in the simulation
(note that they did not include the AGN feedback) and/or environment
might play a role here.  They studied objects that become isolated massive galaxies today, while
we have focused on a forming galaxy group.  A galaxy group is an over-density
region in the universe, where a larger number of density peaks are embedded
in a small volume by definition, and thus group galaxies might experience
an accelerated formation and assembly at early times.
This is an area where further simulations would be useful.

Our observation sets the upper limit on the formation time scale.
The lower limit comes from the observed solar [Fe/H] abundance in
nearby massive galaxies (e.g., \citealt{nelan05}).
The galaxies must have experienced multiple cycles of star formation,
so that the initial primordial gas is enriched to solar metallicity.
In other words, the star formation should not be instantaneous.
Closed-box models have provided a way to link metallicity and 
star formation time scales, but a detailed chemo-hydrodynamic simulation
in the cosmological context would be needed to derive a more useful lower limit.

After the intensive in-situ star formation, subsequent star formation
has to be shut off in order to reproduce the red spectra of the observed
quiescent galaxies without prominent emission lines.
Theoretical work has suggested a role of AGNs for such quenching
\citep{granato04,springel05b,croton06,bower06}.
However, there is still a limited amount of convincing observational evidence
that AGNs actually shut off star formation, except possibly for powerful
QSOs \citep{ho05,kim06,greene11} and powerful radio galaxies \citep{nesvadba08}.
The central radio galaxy of the proto-cluster is also likely powerful
enough to blow out the gas \citep{nesvadba06}.
There are 5 spectroscopically confirmed AGNs in this proto-cluster \citep{croft05}
and 1 out of the 4 quiescent galaxies focused on in this paper is an AGN.
Among all the massive galaxies with $\gtrsim10^{11}M_\odot$, we estimate the AGN
fraction is $\sim 40$\%\footnote{
This should be regarded as a rough number.
We use our $z_{phot}$ and $z_{spec}$ from the literature to select massive
member galaxies here and it is often not trivial to measure stellar mass of
hosts of powerful AGNs.  We assume a constant $M^*/L_{Ks}$ for such objects.
}.
We also obtained a tentative hint that AGNs may populate among
star forming galaxies in Section \ref{sec:stacked_sf}.
This large fraction of galaxies host AGNs in the proto-cluster and
a comparably large AGN fraction is observed in the field environment
at similar redshifts as well (e.g., \citealt{xue10,marchesini10,olsen13}).
Also, a very poor group at $z=1.61$,
which might possibly be a descendant of a forming system like the one studied here,
has a high AGN fraction ($\sim40\%$; \citealt{tanaka13}).
These high AGN fractions hint at a possible role of the AGN feedback
in the quenching of star formation in massive galaxies and could be
considered as smoking-gun evidence.  However, they do not provide direct evidence.
It would be fair to say that a physical mechanism to
shut off star formation is yet to be identified.

Whatever the physical process is, the quenching is likely a fast process.
We have observed a clear separation between the star forming galaxies
and quiescent galaxies in  Fig. \ref{fig:cmd}.
If the quenching happens on a long time scale, we may expect
to observe transition galaxies in between the two populations.
But, such a population does not seem to exist in this proto-cluster.
This fast quenching is in line with 
the short formation time scale inferred
from  our analysis in section 5.

\begin{figure}
\epsscale{1.0}
\plotone{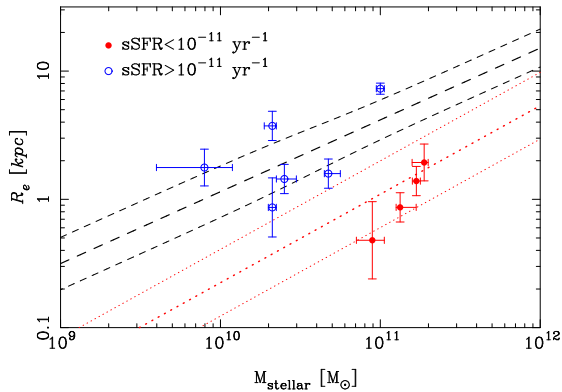}
\caption{
  Effective radius plotted against stellar mass.
  The filled and open circles are quiescent (sSFR$<10^{-11}\ \rm yr^{-1}$)
  and star forming galaxies (sSFR$>10^{-11}\ \rm yr^{-1}$), respectively.
  The dashed lines are the size-mass relation for early-type galaxies 
  at $z=0$  and its $1\sigma$ range from \citet{shen03}.
  The dotted lines show the size-mass relation for quiescent galaxies
  in the field at $2.0<z<2.5$ \citep{newman12}.
  The dotted and dashed lines should be compared with the filled circles.
}
\label{fig:size_mass}
\end{figure}

The intense starburst within $\lesssim0.5$ Gyr
followed by a rapid quenching has to result in compact
early-type galaxies with high S\'{e}rsic indices.
There is now a wealth of observations that massive quiescent
galaxies at $z\sim2$ are compact and have steep radial profiles
(e.g., \citealt{daddi05,toft07,zirm07,cimatti08,vandokkum08,cimatti12};
but see also \citealt{vanderwel11}).
We show in Fig. \ref{fig:size_mass} the size-mass relation for
the proto-cluster galaxies based on the NICMOS
data presented in \citet{zirm08} and \citet{zirm12}.
We find that the proto-cluster quiescent galaxies as shown by the filled
circles are indeed compact compared to early-type galaxies at $z=0$
from \citet{shen03}.
If we compare the proto-cluster quiescent galaxies with field quiescent galaxies
similar redshifts drawn from \citet{newman12},
the proto-cluster quiescent galaxies seem to have similar physical
sizes to those of field galaxies, although the statistics are poor.
\citet{zirm12} reported on possibly larger sizes of the proto-cluster
galaxies compared to the field based on a larger photometric sample,
and it is likely that we suffer from the poor statistics here.

A promising way to produce such a compact object would be nuclear starbursts.
Galaxy mergers induce strong gas inflow towards the center \citep{mihos96},
triggering a nuclear starburst.
Also, filamentary gas inflow, which primarily happens at high redshifts
\citep{keres05,dekel06,johansson12}, can also reach the central region and help feed
the nuclear starburst.

One may expect to observe some of such starbursting galaxies as submm galaxies.
Toft et al. (in prep.) suggested that submm galaxies at $3\lesssim z\lesssim6$
may be the direct progenitors of the quiescent galaxies at $z\sim2$
because the $z_f$ distribution of the quiescent galaxies match
with the observed redshift distribution of submm galaxies.
Some of these submm galaxies have SFRs even above $1000\rm M_\odot\ yr^{-1}$, but
a recent SPH simulation by \citet{johansson12} shows that massive galaxies
reach a maximum SFR of about an order of magnitude lower, $\sim100\rm M_\odot$.
This is a total SFR of all galaxies that end up in a massive galaxy at $z=0$,
and thus SFRs of individual galaxies at $z=2$ should be lower.
This may indicate a tension between observation and simulation or
perhaps it just means that only a fraction of massive galaxies experience
a submm phase (and hence not fully reproduced by simulations of a dozen
galaxies).  We should also mention that the simulated galaxies in
\citet{johansson12} tend to be less massive than observed submm galaxies
and the comparison here may not be fair.
This is also an area that deserves further studies.

The gas continues to accrete to the quenched galaxies from
the surroundings, but they have to remain quiescent in order to be
consistent with the observations at lower redshifts.
Minor mergers help keep the gas hot to prevent further star formation
through the release of the gravitational energy of the infalling galaxies \citep{johansson09}.
This heating may not be enough and the gas may still cool, especially in the core \citep{johansson12}.
The so-called radio mode AGN feedback might be able to heat it
\citep{croton06,bower06}, although this process is not necessarily
confirmed by observation.
Minor mergers also help explain the observed size evolution of massive
quiescent galaxies (e.g., \citealt{naab09,oser12}) down to the present day.

Overall, we have not fully understood the physical processes that trigger
such star formation, enrich metals to solar metallicity, produce a compact
galaxy, and then shut off SFR and keep them quiescent afterwards,
but aided by numerical simulations, we are starting to constrain them.
Our  observational constraint presented in this paper is that
the massive quiescent galaxies form at $3\lesssim z_f\lesssim4$ on
a $\lesssim0.5$ Gyr time scale.
This short formation time scale is in tension with the recent models as 
we have highlighted in this paper.
But, this might indicate that environment plays a role --- simulations
have focused on field galaxies, but we specifically studied
a forming group, and it would not be too surprising if their formation
time scales are different.
From an observational perspective, it would be interesting to obtain
higher S/N spectra as well as to study
a larger number of such systems at high redshifts.
From a theoretical perspective, a detailed simulation of galaxy formation
in high density regions of the universe would be highly useful.

\bigskip

This work is based on data collected at Subaru Telescope, which is
operated by the National Astronomical Observatory of Japan. 
MT acknowledges support by
KAKENHI No. 23740144 and thanks Peter Johansson for useful discussions
and comments on the manuscript.
MT also thanks Kiyoto Yabe for useful conversations.
ST and AZ gratefully acknowledge support from the Lundbeck foundation.
The Dark Cosmology Centre is funded by the Danish National Research Foundation.
DM acknowledges support from Tufts University Mellon Research Fellowship in Arts and Sciences.
YK acknowledges the support from the Japan Society for the Promotion of Science
(JSPS) through JSPS research fellowships for young scientists.
We thank the referee for useful comments, which helped improve the paper.

\appendix

\section{Spectra of objects with measured redshifts}

\begin{figure}
\epsscale{1.}
\plotone{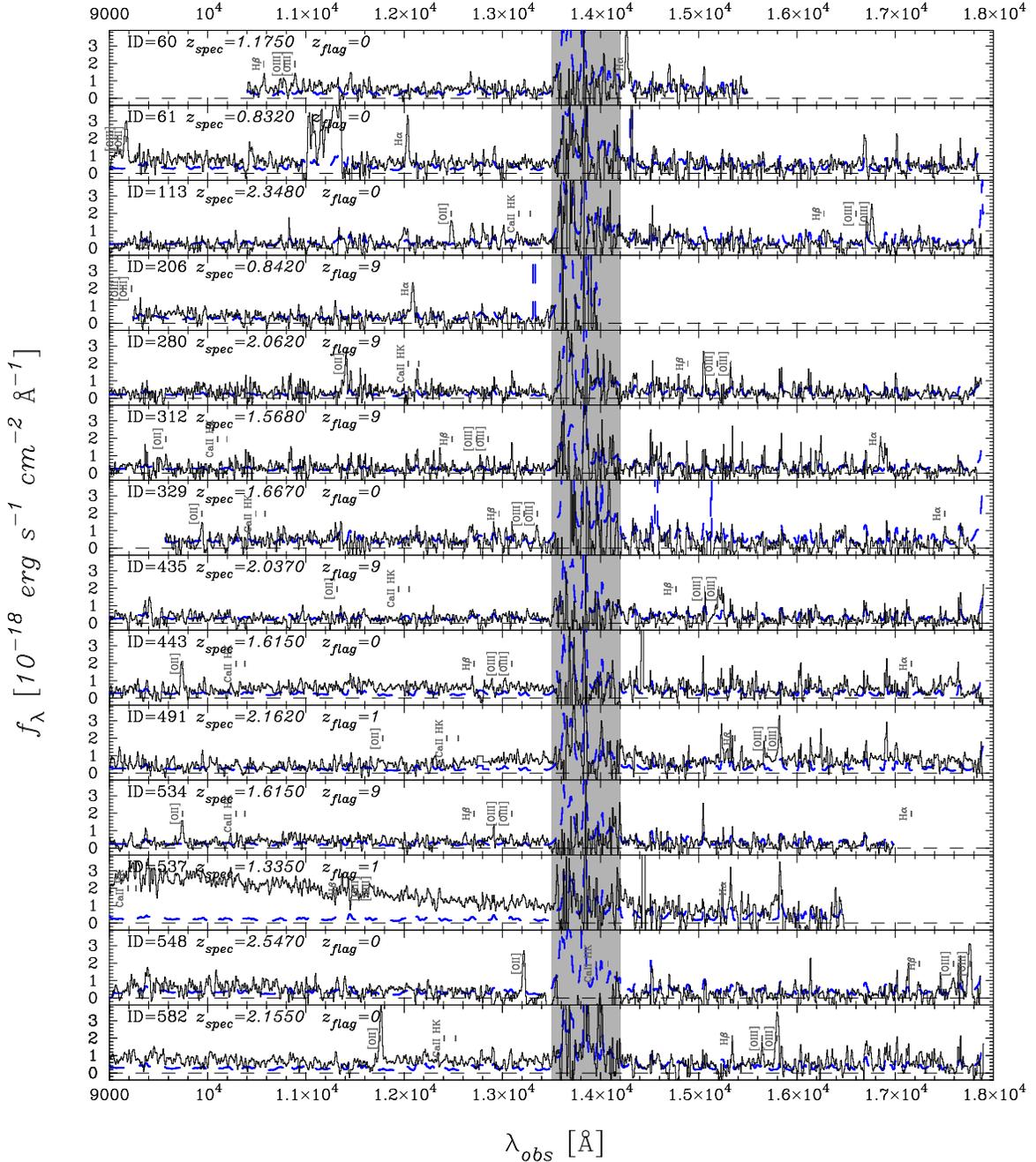}
\caption{
  Spectra of objects with measured redshifts in Table \ref{tab:spec_sample}.
  The object and noise spectra are shown by the solid and dashed lines, respectively.
  We apply a boxcar smoothing of width $20\rm\AA$ to both of them.
  The object ID, redshift, and confidence flag are indicated in each panel.
  The flags mean: 0=secure, 1=possible, 9=single emission line.
  The shaded area is strongly affected by the atmospheric absorption.
}
\label{fig:specall}
\end{figure}

We present the spectra of objects with measured redshifts in Fig. \ref{fig:specall}.
As shown in Table \ref{tab:spec_sample}, we could not measure spectroscopic redshifts
for roughly a half of the objects that we observed with MOIRCS.  Those spectra
are not included in the figure.

\section{Improvements on redshift and physical parameter estimates with the binned spectra}

\begin{figure}
\epsscale{0.5}
\plotone{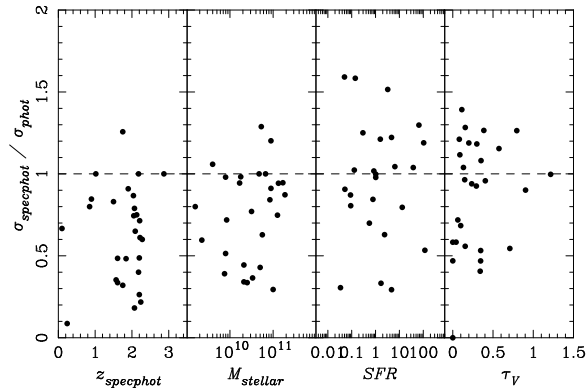}
\caption{
  Improvements on redshift and physical parameter estimates.
  The vertical axis shows the ratio of the uncertainties with
  and without the spectra.  The panels are for redshift,
  stellar mass, SFR, and $\tau_V$ from the left to the right.
}
\label{fig:spec_yesno}
\end{figure}

It will be instructive to show how our estimates of redshifts and physical
parameters of galaxies improve by combining the binned spectra
in the SED fits.  We plot in Fig. \ref{fig:spec_yesno} the ratio of
the 68\% confidence intervals on redshift, stellar mass and SFR
measured with and without the binned spectra.  If the ratio is below
unity, that means that the binned spectra reduce the uncertainty.

Before we discuss the plot, it is important to emphasize that
the result here should not be interpreted as a general improvement
that can be achieved by including spectra in SED fits.  Improvements
are dependent on wavelength coverages of spectra and broad-band
photometry and also on their quality.  Our spectra cover only
a limited rest-frame wavelength range and a wider wavelength
coverage is expected to deliver further improvements.

The left panel shows that the redshift uncertainty is reduced by almost
a factor of 2 on average by the spectra.
There are not many galaxies at low redshifts, but
we expect that this improvement is limited to $1.3\lesssim z\lesssim3.2$,
where we can probe the 4000\AA\ break with the spectra.
At lower/higher redshift ranges, the spectra sample only smooth continua
and they  will not be as useful as the break feature.
This plot clearly shows that our work significantly benefited from
the spectra in identifying probable cluster member candidates
at $z=2.16$.

The strength of the 4000\AA\ break feature is a reasonable proxy for
stellar mass to light ratio (e.g., \citealt{kauffmann03}).  The precise
measurement of the break strength with the spectra improves the stellar
mass estimates as shown in the 2nd panel.  
But, interestingly, the spectra do not improve the SFR estimates (3rd panel).
This is likely because the wavelength coverage of the spectra is too narrow
and most of the constraints on the overall spectral shape come from
the broad-band photometry.  Not surprisingly, we find that the spectra
do not improve dust extinction estimates either (4th panel).
Rest-frame UV spectra will be  useful to improve SFR and extinction.


\end{document}